\documentclass[10pt]{article} 
\usepackage[accepted]{tmlr}

\usepackage{amsmath,amsfonts,bm}

\def\eqref#1{equation~\ref{#1}}

\def\1{\bm{1}}

\DeclareMathAlphabet{\mathsfit}{\encodingdefault}{\sfdefault}{m}{sl}
\SetMathAlphabet{\mathsfit}{bold}{\encodingdefault}{\sfdefault}{bx}{n}

\usepackage{hyperref}
\usepackage{url}
\usepackage{booktabs}           
\usepackage{multirow}           
\usepackage{amsfonts}           
\usepackage{graphicx}           
\usepackage{duckuments}         

\usepackage{amsmath,amssymb} 
\usepackage{xspace}

\usepackage[capitalize]{cleveref}
\usepackage{caption}
\usepackage{subcaption}
\usepackage{graphicx}  
\usepackage{tabularx}
\newcolumntype{Y}{>{\centering\arraybackslash}X}
\usepackage{array,multirow}
\usepackage{wrapfig}
\usepackage{colortbl}

\usepackage{svg}
\usepackage{comment}
\usepackage{soul}

\usepackage{pifont}
\newcommand{\cmark}{\ding{51}}
\newcommand{\xmark}{\ding{55}}
\newcommand{\DCG}{\text{DCG}}

\newcommand{\NDCG}{\text{NDCG}}

\newcommand{\LNDCGs}{\mathcal{L}_{\text{NDCG}_s}}
\newcommand{\AP}{\text{AP}}

\newcommand{\LAPs}{\mathcal{L}_{\text{AP}_s}}

\newcommand{\LRks}{\mathcal{L}_{\text{R}@k_s}}
\newcommand{\etal}{\textit{et al}. }

\DeclareUnicodeCharacter{0301}{\'{e}}
\newcommand{\ie}{\emph{i.e.}\xspace}
\newcommand{\eg}{\emph{e.g.}\xspace}

\DeclareMathOperator{\rank}{rank}

\newcommand{\llargo}{\mathcal{L}_{\text{ITEM}}}

\title{ITEM: Improving Training and Evaluation of Message-Passing based GNNs for top-$k$ recommendation}

\author{\name Yannis Karmim \email yannis.karmim@cnam.fr \\
      \addr Conservatoire national des arts et métiers, CEDRIC, Paris, France \\
      \AND
      \name Elias Ramzi \email elias.ramzi@cnam.fr \\
      \addr Conservatoire national des arts et métiers, CEDRIC, Paris, France\\
      \AND
      \name Raphaël Fournier S'niehotta \email fournier@cnam.fr \\
      \addr Conservatoire national des arts et métiers, CEDRIC, Paris, France
      \AND
      \name Nicolas Thome \email nicolas.thome@isir.upmc.fr \\
      \addr Sorbonne Université, CNRS, ISIR, Paris, France\\
      \\
      }

\def\month{07}  
\def\year{2024} 
\def\openreview{\url{https://openreview.net/forum?id=9B6LM2uoEs}} 

\begin{document}

\maketitle

\begin{abstract}
Graph Neural Networks (GNNs), especially message-passing-based models, have become prominent in top-$k$ recommendation tasks, outperforming matrix factorization models due to their ability to efficiently aggregate information from a broader context. 
Although GNNs are evaluated with ranking-based metrics, \eg \textsc{NDCG}@k and Recall@k, they remain largely trained with proxy losses, \eg the \textsc{BPR} loss. In this work we explore the use of ranking loss functions to directly optimize the evaluation metrics, an area not extensively investigated in the GNN community for collaborative filtering.
We take advantage of smooth approximations of the rank to facilitate end-to-end training of GNNs and propose a Personalized PageRank-based negative sampling strategy tailored for ranking loss functions. Moreover, we extend the evaluation of GNN models for top-$k$ recommendation tasks with an inductive user-centric protocol, providing a more accurate reflection of real-world applications.
Our proposed method significantly outperforms the standard \textsc{BPR} loss and more advanced losses across four datasets and four recent GNN architectures while also exhibiting faster training. Demonstrating the potential of ranking loss functions in improving GNN training for collaborative filtering tasks.
\end{abstract}

\section{Introduction}\label{sec:introduction}

Recommender systems have become an essential component in many online applications, helping users discover relevant and personalized content amid the overwhelming abundance of information available on the internet. Collaborative filtering is one of the most popular and widely adopted techniques for building recommender systems, which operates by leveraging the past behavior of users and the relationships among items to generate recommendations.
\begin{figure}[!hbt]
    \centering
    \includegraphics[width=\textwidth]{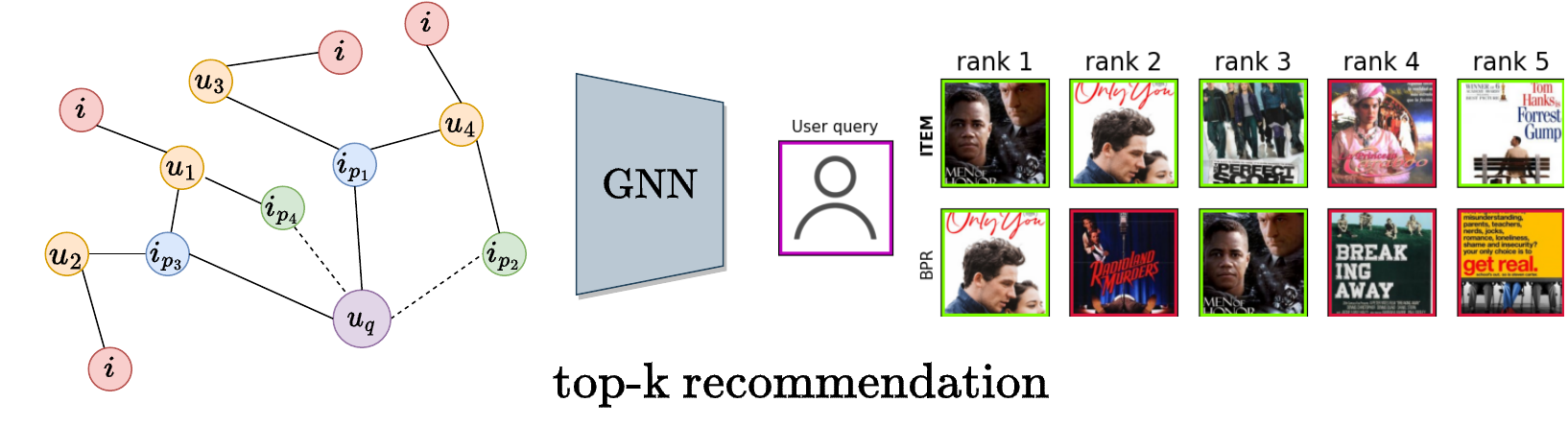}
    \caption{
    The goal in top-$k$ recommendation is to recommend to a user, \eg $u_q$ (purple), relevant items such as $i_{p_4}$ (in green), based on its interaction history, \ie items in blue such as $i_{p_1}$. \textsc{ITEM} directly optimizes the evaluation metric, \ie \textsc{NDCG}, during training using a smooth approximation of the rank and Personalized PageRank (\cite{Page1998TheWeb}) based negative sampling.  Best seen in color.}
    
    \label{fig:figure_intro}
\end{figure}
Graph Neural Network architectures have emerged as powerful methods for learning and representing complex data structures, particularly those that exhibit non-Euclidean properties such as graphs (\cite{Velickovic2017GraphNetworks,Hamilton2017InductiveGraphs,Kipf2016Semi-SupervisedNetworks,Rossi2020TGN:Graphs,Xu2018HowNetworks}).
In the context of recommender systems, GNNs offer a natural way to model user-item interactions, leveraging the inherent graph structure to capture higher-order relationships among users and items. Among the various GNN architectures, message-passing GNNs (MP-GNNs) have demonstrated impressive performance on top-$k$ recommendation tasks (\cite{He2020LightGCN:Recommendation,Wang2019NeuralFiltering,Sun2020NeighborRecommendation,Chen2020RevisitingApproach}), consistently surpassing traditional matrix factorization methods (\cite{Rendle2012BPR:Feedback,Hsieh2017CollaborativeLearning,Chen2020EfficientRecommendation}).

Top-$k$ recommendation is a standard task in recommender systems, which involves generating an ordered list of k items for a user. This process necessitates calculating similarities between elements before ranking them. Evaluation of top-$k$ recommendations typically employs ranking-based metrics such as Average Precision (AP), Normalized Discounted Cumulative Gain (\textsc{NDCG}), or Recall at k (R@k). However, these metrics are non-differentiable and cannot be directly used to train neural networks like GNNs with Stochastic Gradient Descent (SGD).

\vspace{0.3cm}
Instead, GNNs are commonly trained using a pairwise loss, the well-known Bayesian Personalized Ranking (\textsc{BPR})  (\cite{Rendle2012BPR:Feedback}), which serves as a coarse approximation of the ranking metric. The BPR is not explicitly designed to optimize standard non-differentiable evaluation rank-based metrics, such as \textsc{NDCG}@K and Recall@K:
there are some criticisms about the gap between the evaluation objective and the training objective of the BPR (\cite{Wu2021Self-supervisedRecommendation}).
We investigate alternative loss functions that more closely align the training and evaluation objectives.

The second issue arises from another misalignment, between batch learning and rank-based evaluation.  During training, one does not have access to the true rank and must resort to hard negatives, that are more challenging to distinguish from positive items, for better performance.
Some methods, such as MixGCF (\cite{Huang2021MixGCF:Systems}), generate artificial hard negatives for the BPR loss. While this approach is suitable for a pairwise loss like BPR, its online process is computationally expensive. Given that we employ a listwise loss function, we need to sample a large number of informative negative items per user, the hard negative generation of MixGCF is not tractable.  We thus propose to sample negative items offline, based on their Personalized PageRank (PPR) score (\cite{Page1998TheWeb}): a high PPR score indicates proximity in the graph, making them more challenging to distinguish from a positive item for MP-GNNs.

Another impediment in the GNN literature for top-$k$ recommendation is the limited and unrealistic evaluation protocol. Recent works predominantly use a transductive approach, evaluating on the same users used in training. This not only deviates from real-world recommendation scenarios but also fails to consider GNN models' generalization capacity. We propose to enhance evaluation by incorporating an inductive user-split protocol, evaluating models on users not seen during training.

In this paper, we introduce our framework \textsc{ITEM} (\textbf{I}mproving \textbf{T}raining and \textbf{E}valuation of \textbf{M}essage-passing-based GNNs). Our framework is designed to provide training and evaluation for Graph Neural Networks tailored to the top-$k$ recommendation task.
\begin{itemize}
    \item Specifically, our list-wise loss $\llargo$ first leverages smooth rank approximations, which have recently been revisited in image retrieval and machine learning~\cite{bruch2019revisiting,brown2020smooth,ramzi2021robust},
~leading to good approximations of the evaluation metrics, such as the \textsc{NDCG} or AP.
    \item Additionally, we enhance our loss $\llargo$  by incorporating a negative sampling strategy tailored for rank approximation losses and leveraging the graph data structure. This strategy is based on the Personalized PageRank (\cite{Page1998TheWeb}) (PPR) score. We show that this sampling is particularly well suited to our loss $\llargo$, since it allows for a fast sampling of many informative negative items. This sampling helps to build large efficient batches to better approximate the true ranking. 

    \item Finally, we propose to evaluate and benchmark GNNs for top-$k$ recommendation in an inductive user-split protocol. While it is known in the field of recommendation and used by some traditional models (\cite{Meng2020ExploringModels,Liang2018VariationalFiltering}), this setting has not been used to evaluate GNN architectures. This user-split protocol is more realistic because it introduces new users in testing, thus better evaluating the generalization capacity of recommender systems. 

\end{itemize}

We carry out extensive experimental validations in both 
transductive and inductive settings. Our results highlight the benefits of \textsc{ITEM} over the standard training with \textsc{BPR} loss in terms of time and performance across multiple GNN architectures. Moreover, we demonstrate that \textsc{ITEM} outperforms more advanced state-of-the-art loss functions, showcasing its effectiveness.

\section{Related Work}\label{sec:related_work}

\subsection{Graph Neural Networks for Collaborative Filtering} We focus on collaborative filtering (CF) models applied to data with implicit feedback, where only connections between users and items are considered, without incorporating other informations such as rating. A common variant of CF is the top-$k$ recommendation task, where the goal is to identify a small set of items that are most relevant to a user's interests. In this context, Graph Neural Networks have shown impressive results, indeed they are directly suitable to this task since user-item interactions can be modeled by bipartite graphs. 
Within the family of graph neural network models, message-passing-based (\cite{Gilmer2017NeuralChemistry}) methods  have demonstrated superior performance compared to traditional CF models. These include matrix factorization (MF) (\cite{Rendle2012BPR:Feedback,Hsieh2017CollaborativeLearning,Chen2020EfficientRecommendation}), auto-encoders (\cite{Liang2018VariationalFiltering}), and node embedding models that rely on random walks for generating representations (\cite{Perozzi2014DeepWalk:Representations,Grover2016Node2vec:Networks}).
MP-GNNs for CF learn user and item representations by propagating and updating their embeddings through the bipartite graph. The \textsc{BPR} loss is employed to optimize the user-item representations by ensuring that a user's embedding is closer to a positive item than to a negative one. In this sense several models of MP-GNNs have been designed such as NGCF (\cite{Wang2019NeuralFiltering}), LR-GCF (\cite{Chen2020RevisitingApproach}), or DGCF (\cite{Wang2020DisentangledFiltering}).\\
He \etal introduce LightGCN (\cite{He2020LightGCN:Recommendation}) as a simplified version of NGCF, achieved by eliminating weight matrices and non-linear activation layers. Although LightGCN is less expressive, it proves to be highly effective and more efficient, with a considerably simplified training process. Subsequent research has focused on enhancing the training of these models: SGL-ED (\cite{Wu2021Self-supervisedRecommendation}) proposes to combine the standard \textsc{BPR} loss with a self supervised loss, while MixGCF (\cite{Huang2021MixGCF:Systems}) artificially generates hard negatives embeddings for negative sampling to replace the random negative sampling in the standard \textsc{BPR} loss. 
\subsection{Evaluation protocol for top-$k$ recommendation}
All these GNN models competed with matrix factorization methods thus, their learning and evaluation setups are transductive.

However, other evaluations and training protocols for recommendation exist (\cite{Liang2018VariationalFiltering}). In (\cite{Meng2020ExploringModels}), authors show that the evaluations of GNNs-based models are limited compared to all those existing in the literature. Instead of splitting the data based on interactions, one may split the data by user, with some users in the training set, and new users for testing (see~\cref{fig:figure_protocol}). This setup is closer to real applications, and allows to produce recommendation to new users without re-training the model. It is also more challenging, requiring to construct representations for a new user without learning. 

We propose here to evaluate our \textsc{ITEM} model and benchmark MP-GNNs baselines in this realistic setting which was not used before for MP-GNNs, to the best of our knowledge.

\subsection{Ranking-based loss function}

The Learning to Rank problem (\cite{10.1145/1273496.1273513}) is evaluated using ranking based metrics, \eg \textsc{NDCG} (\cite{jarvelin2002cumulated}), R@K, or AP (\cite{croft2010search}).
~As these metrics are not differentiable (because of the ranking operator), their optimization has been abundantly studied. Different proxy methods have been built in Information Retrieval, with for instance pairwise loss (\cite{ranknet,hadsell2006dimensionality}), or triplet losses (\cite{Rendle2012BPR:Feedback,li2017joint}). To train GNNs, the most well-known loss is the \textsc{BPR} loss (\cite{Rendle2012BPR:Feedback}), a smoothed triplet loss. However, it was shown that triplet losses tend to put more emphasis on correcting errors at the bottom of the ranked list, rather than at the top, which would have the most impact to maximize the metric (\cite{brown2020smooth}).
The {\em direct} optimization of ranking-based metrics has long been studied for Information Retrieval using structured SVM (\cite{Yue:2007}), or rank approximations (\cite{lambdarank,softrank,Qin2009AGA}). Several direct optimization methods have gained traction in deep learning using for instance soft binning approaches (\cite{revaud2019learning}) or rank approximations (\cite{bruch2019revisiting,https://doi.org/10.48550/arxiv.2102.07831,brown2020smooth,ramzi2021robust,ramzi2022hierarchical}). \textcolor{black}{ NeuralNDCG \cite{https://doi.org/10.48550/arxiv.2102.07831} proposed a differentiable approximation to NDCG by using NeuralSort, a differentiable relaxation of the sorting operator, resulting in a smooth variant of the metric and a new ranking loss function. ROADMAP \cite{ramzi2021robust} introduce a robust and decomposable average precision (AP) loss for image retrieval, addressing non-differentiability and non-decomposability issues with a new rank approximation and calibration loss. \cite{ramzi2022hierarchical} use a hierarchical AP training method for pertinent image retrieval, leveraging a concept hierarchy to refine AP by integrating errors’ importance and better evaluate rankings. In our work, we introduce the use of the sigmoid function for approximating the NDCG rank~\cite{brown2020smooth,Qin2009AGA} on GNNs in a top-$k$ recommendation context.}

\section{ITEM framework}\label{sec:method}

In this section, we present the \textsc{ITEM} framework. We first define our ranking-based loss using a smooth approximation of the rank in~\cref{sec:metric_optim} as well as our adapted  negative sampling strategy in~\cref{sec:neg_sample} . We then introduce the protocol used to train and evaluate the GNNs performances in~\cref{sec:evaluation}.

\begin{figure}
    \centering
    \includegraphics[width=\textwidth]{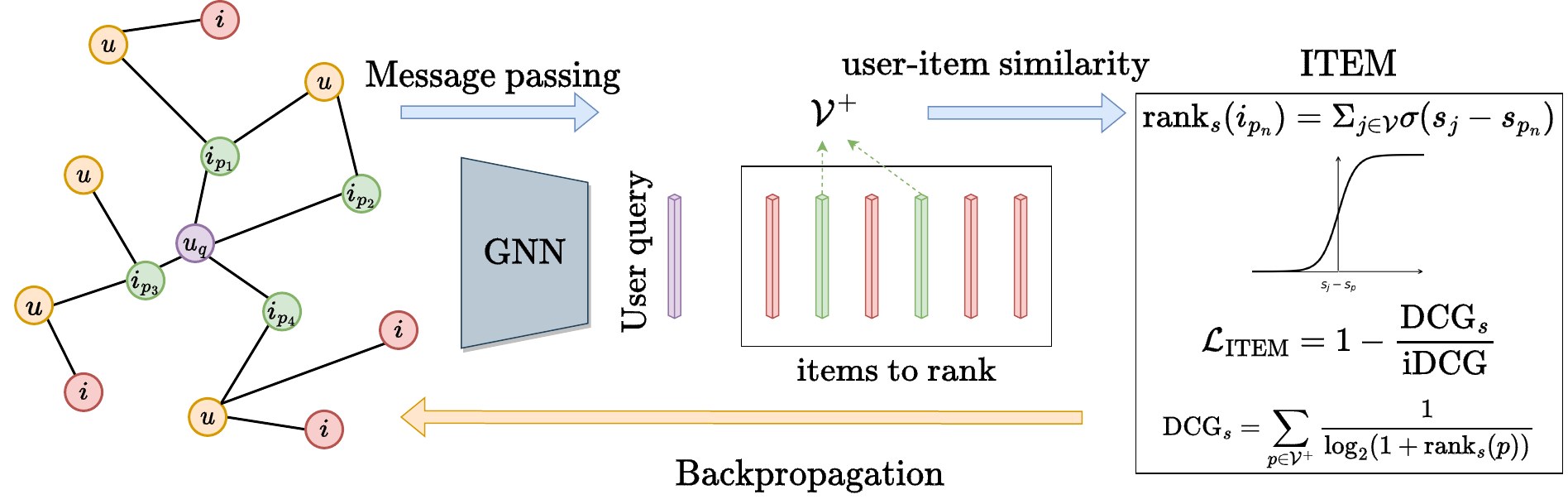}
    \caption{Using message passing, a GNN creates embeddings for every node of the graph. For each user we first construct a batch of randomly sampled positive items and negative items selected with our Personalized PageRank (\cite{Page1998TheWeb}) based negative sampling~\cref{eq:pprneg}. We then compute the score of the user wrt. the batch of items and calculate the loss using the approximation of the rank of~\cref{eq:def_approx_rank}. Finally the loss is backpropagated to update the parameters of the GNN, and update the embeddings for the items and users in the transductive setting. Best seen in color.}
    \label{fig:figure_method}
\end{figure}

\textbf{Training context} We consider an undirected bipartite graph $\mathcal{G}=(\mathcal{U},\mathcal{V},\mathcal{E})$ with $|\mathcal{U}|$ users, $|\mathcal{V}|$ items and $|\mathcal{E}|$ edges. We assign to each node in $\mathcal{U} \cup \mathcal{V}$ an embedding $\mathbf{h}$. We use a GNN that re-embeds the node embeddings to another space of the same dimension using message passing. 

The task is to construct an embedding space such that, after message passing, the embedding of a user is closer to the embeddings of its positive items ($\mathcal{V}^+$) than to its negative items ($\mathcal{V}^-$), \ie for user $u$, its embedding $\mathbf{h_u}$, a positive item $p$, its embedding $\mathbf{h_p}$, and a negative item $j$, we want $s_p > s_j, \text{with } s_p = \mathbf{h_u}\cdot \mathbf{h_p}^T, \text{ and } s_j=\mathbf{h_u}\cdot \mathbf{h_j}^T$, $s_p$ and $s_j$ are respectively the similarity score between the positive and the negative item. To evaluate the performances of a GNN, we use ranking-based metrics, \textsc{NDCG}~\cref{eq:def_ndcg_metric}, that measures the quality of a ranking. 

\begin{equation}
    \NDCG = \frac{\DCG}{\text{iDCG}}, \text{ with } 
    \begin{cases}
    \DCG = \sum_{p\in\mathcal{V}^+} \frac{1}{\log_2(1+\rank(p))} \\
    \text{iDCG} = \max_{\text{ranking}} \DCG
    \end{cases}
     \label{eq:def_ndcg_metric} 
\end{equation}

\subsection{Direct ranked-based optimization}\label{sec:metric_optim}

GNNs are evaluated using standard ranking based metrics, \eg \textsc{NDCG}, R@k, AP. We propose to train GNNs by optimizing directly smooth approximations of those metrics. Specifically, we use an approximation of the ranking operator to yield losses amenable to gradient descent.

The ranking operator can be defined as: $\rank(p) = 1 + \sum_{j\in\mathcal{V}} H(s_j - s_p)$, with $H$ the Heaviside (step) function (\cite{Qin2009AGA,brown2020smooth}). The intuition behind this definition is that in order to have the rank of a (positive) item $p$, we must ``count'' the number of items~$j$ that have a similarity to the query user $s_j$ greater than $p$'s, $s_p$, \ie $H(s_j - s_p)=1$. During training, we aim to minimize ${\rank^- = 1 + \sum_{j\in\mathbf{\mathcal{V}^-}} H(s_j - s_p)}$, \ie the number of ``negative'' items that have a higher score than positive items. Ranking-based metrics optimize this objective.
This writing of the rank shows why the ranking operator is not differentiable, \ie because the Heaviside function is not, specifically its gradients are null or undefined.

We propose to use the sigmoid function as an approximation of the Heaviside function (\cite{Qin2009AGA,brown2020smooth}): $\sigma(x;\; \tau) = \frac{1}{1 + \exp^{\frac{-x}{\tau}}}$, with $\tau\in\mathbb{R}$ a temperature scaling parameter. $\tau$ controls the slop of the sigmoid, as $\tau$ gets smaller the slope is greater, and the sigmoid saturates faster.

Using this approximation we can define a smooth version of the rank:

\begin{equation}\label{eq:def_approx_rank}
    \rank_s(p, \tau) = 1 + \sum_{j\in\mathcal{V}} \sigma(s_j - s_p;\; \tau)
\end{equation}
$\rank_s$ is differentiable and is thus amenable to gradient descent. It has a single hyper-parameter, $\tau$, we study its impact in our experimental validation (\cref{sec:model_analysis}).

\textbf{Application to ranked-based metrics} This plug-and-play rank approximation can be used to get a smooth version of ranking metrics, \eg \textsc{NDCG}, R@k and AP.

\textcolor{black}{The} approximation of the Normalized Discounted Cumulative Gain (\textsc{NDCG}) is defined as follows:
\begin{equation}\label{eq:def_ndcg}
    \llargo = 1 - \frac{\DCG_s}{\text{iDCG}}, \text{ with } \DCG_s = \sum_{p\in\mathcal{V}^+} \frac{1}{\log_2(1+\rank_s(p))}
\end{equation}
We use \textcolor{black}{this} approximation of $\rank_s$ to approximate the DCG (see~\cref{eq:def_ndcg_metric}), and use the exact iDCG (see~\cref{eq:def_ndcg_metric}). \textcolor{black}{In Sec. A of supplementary material we show how other ranking-based metrics can be approximated and show their effect on the evaluation metrics in Sec. B.5.}

\textbf{End-to-end ranked-based training of GNN} GNNs, \eg GCN (\cite{Kipf2016Semi-SupervisedNetworks}), jointly learn embeddings for the items and users of a graph, and through parameterized (or not for LightGCN \cite{He2020LightGCN:Recommendation}) message passing they create representation for items and users. Using a user as a query (\textcolor{black}{purple} embedding in~\cref{fig:figure_method}), as in Learning to Rank (\cite{10.1145/1273496.1273513}), we aim to make the distance between a user's embeddings and the ones of its positive items, \ie with implicit feedback (in \textcolor{black}{green} on~\cref{fig:figure_method}), closer than the negatives ones, \ie no implicit feedback (in \textcolor{black}{red} on~\cref{fig:figure_method}). \textsc{ITEM}~\cref{eq:def_ndcg} is directly applied on the similarities, to produce better ranking. As \textsc{ITEM} is differentiable, after computing the loss we can backpropagate gradients through the network (see~\cref{fig:figure_method}), to update the potential weights of the GNNs, and update the users and items embeddings. 

\subsection{Negative Sampling}\label{sec:neg_sample}The objective of \textsc{ITEM} is to rank, for a user query, its positive items (high rating or implicit feedback) before the negative ones. 
To do so, for each user we have to construct, a batch of positive items and negative items as we cannot use all items. For a given user the number of negative items is much larger than that of the positive items. In order to better approximate the global ranking in mini-batch learning, we have to sample a significant number of informative negative items. Uniform random sampling is adopted as a solution in many GNN models for recommendation (\cite{He2020LightGCN:Recommendation,Wang2019NeuralFiltering}), however the sampled negatives are often not very informative which can limit the performances of the model. 
We propose to use the Personalized PageRank (\cite{Page1998TheWeb}) (PPR) to weigh the sampling of negative examples. Specifically we normalize the PPR score using the softmax function and sample negative items $j$ for a user query $u_q$ according to~\cref{eq:pprneg}. Indeed items with high PPR score will be harder to rank, as they are in closer proximity to the user query.
\begin{equation}
    \label{eq:pprneg}
    p({i}^{-}_{j}|u_q)\quad \sim \quad \frac{e^{\text{ppr}_{u_q}({i}^{-}_{j})}}{\sum_{k\in\mathcal{V}^-}e^{\text{ppr}_{u_q}({i}^{-}_{k})}}
\end{equation}
Unlike PinSage (\cite{Ying2018GraphPinSage}), our sampling strategy evaluates all negative items by weighing them differently. \textcolor{black}{Specifically, the PPR score is calculated for each target user, assessing its proximity to every item in the bipartite graph. These scores are normalized with the softmax function to create a probability distribution for sampling. In contrast to existing sampling strategies such as MixGCF (\cite{Huang2021MixGCF:Systems}) which generate hard negatives from embeddings "online" (i.e., during training), our PPR score can be computed on the bipartite graph "offline", before training. The offline computation makes hard negative sampling much faster, allowing us to build informative and large batches well-suited for our $\llargo$ loss.
We perform ablative experiments in~\cref{tab:sampling_compa} to show the benefit of this sampling strategy. 
Our strategy efficiently selects informative hard negative items, but runs the risk of including false negatives. Alternative strategies may avoid false negatives entirely, but result in non-informative samples. 
Our results in \cref{tab:sampling_compa} demonstrate that our trade-off, which samples hard negatives, in spite of the risk of false negatives, improves model performances.}

\subsection{Protocol for collaborative filtering}\label{sec:evaluation}

\begin{figure}
    \centering
    \includegraphics[width=\textwidth]{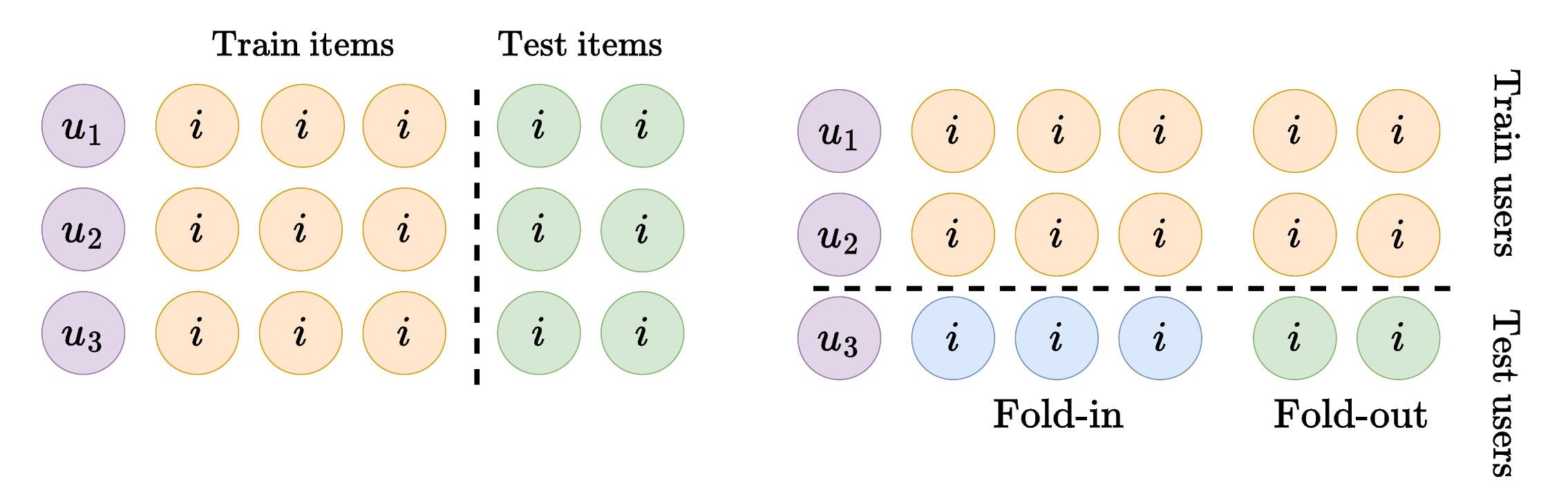}
    \caption{Transductive interaction-split (left) vs Inductive user-split (right) (\cite{Meng2020ExploringModels}). In the first case, the same users are in train and test, their learned embeddings can be directly used in test. In the second case, a part of the users and 100\% of their interactions are used in train. During the evaluation, the model infers a representation of a new test user from some interactions (fold-in), in order to predict the fold-out items where we apply the ranking metrics.}
    \label{fig:figure_protocol}
\end{figure}

\subsubsection{Inductive protocol} We propose to evaluate GNNs in the top-$k$ recommending task  using an inductive user-split protocol (\cite{Meng2020ExploringModels,Liang2018VariationalFiltering}) (illustrated on the right side of~\cref{fig:figure_protocol}). During training, we have access to a fraction $\mu$ (\ie, $u_1$, $u_2$ on~\cref{fig:figure_protocol}) of the user and their \textit{entire} click history, for evaluation we consider the $1-\mu$ (\ie $u_3$ on~\cref{fig:figure_protocol}) unseen users and, with access to a fraction $\eta$ of their click history (\textit{fold-in} -- in blue on~\cref{fig:figure_protocol}), we must recommend the other $1-\eta$ fraction (\textit{fold-out} -- in green on~\cref{fig:figure_protocol}) of their click history. \textcolor{black}{Despite the benefits of this inductive protocol known in top-$k$ recommendation, to our knowledge no GNN models are measured in this user-split protocol in this task.} This protocol measures the capacity of a GNN to make accurate recommendation for a new user (not seen during training), based on its first interactions.

\subsubsection{Transductive protocol} The standard protocol for evaluating GNNs in a recommendation task (\cite{Wang2019NeuralFiltering,He2020LightGCN:Recommendation,Wu2021Self-supervisedRecommendation}) is to evaluate them in a transductive manner, \ie items \textit{and} users are the same during training and evaluation (left of~\cref{fig:figure_protocol}). The training and evaluation splits are defined using an interaction-split, specifically for a given user we use a fraction $\rho$ (orange on~\cref{fig:figure_protocol}) of its interaction to perform message passing during training and for evaluation the GNN must give a higher score to the $1-\rho$ (green on~\cref{fig:figure_protocol}) fraction of its other interactions before the items to which it has no interaction with. For this protocol, the nodes in the graph remain the same for training and evaluation: the system cannot accommodate new users, it is limited to presenting new content to existing users. 

\section{Experiments}\label{sec:experiments}

In our experiments and ablation studies, our aim is to demonstrate the effectiveness of our ranking loss function as well as our negative sampling setup applied to MP-GNNs in comparison to the standard \textsc{BPR} loss and other training enhancement strategies such as self-supervised regularization loss or alternative negative sampling techniques. Additionally, we propose to refine the evaluation of GNNs for CF by benchmarking them using an inductive, user-centric protocol.

\subsection{Datasets and evaluation protocol}
\subsubsection{Evaluation metrics}
We follow (\cite{Huang2021MixGCF:Systems,He2020LightGCN:Recommendation,Wang2019NeuralFiltering}) evaluation protocol, computing the \textsc{NDCG}@20 and Recall@20 with the all-ranking protocol (\cite{He2017NeuralFiltering}).

\begin{table}
    \caption{Statistics of the four datasets used for our experimental validation.}
    \label{tab:datasets_statistics} 
    \setlength\tabcolsep{0pt}
    \begin{tabularx}{1\textwidth}{ l YYY}
        \toprule
         Dataset & \#Users & \#Items & \#Links \\
         \midrule
         MovieLens-100K (\cite{harper2015movielens}) & 610 & 8957 & 100k \\
         MovieLens-1M (\cite{harper2015movielens}) & 6022 & 3043 & 1m \\
         Yelp-2018 (\cite{asghar2016yelp}) & 31668 & 38048 & 1.5m \\
         Amazon-book (\cite{Wang2019NeuralFiltering}) & 52643 & 91599 & 2.9m \\
         \bottomrule
    \end{tabularx}
\end{table}

\subsubsection{Datasets}
We validate our method on four recommendation datasets. We use the MovieLens-100k and MovieLens-1M (\cite{harper2015movielens}) dataset, the 2018 edition of the dataset from the Yelp challenge (\cite{asghar2016yelp}) and Amazon-book (\cite{Wang2019NeuralFiltering}). We detail the preprocessing of the datasets in Sec. B.1 in supplementary.

\subsubsection{Data splitting strategies}

\textbf{Transductive setting:}
For the transductive setup we use the exact same data preprocessing and split as the LightGCN (\cite{He2020LightGCN:Recommendation}) and MixGCF (\cite{Huang2021MixGCF:Systems}) models. In the transductive split, for each user 80\% of these interactions have been randomly selected for the training set and the 20\% for the test and validation set. We followed this same ratio across all datasets in our experiments. This data split is illustrated on the left side of~\cref{fig:figure_protocol}. \newline 

\textbf{Inductive setting:} The inductive setup is based on the user-split data protocol, detailed in \cite{Meng2020ExploringModels} and used by the variational encoder model Mult-VAE (\cite{Liang2018VariationalFiltering}). We used the same ratio as in the Mult-VAE paper, \ie In this protocol, we first separate users into training,validation and test sets. 80\% of the users are kept for training, 10\% for validation and 10\% for testing. Unlike the test and validation sets, the training users keep 100\% of their interaction history for learning our models. The test and validation user interactions are separated into a \textit{fold-in} and a \textit{fold-out} set. The \textit{fold-in} set is used to build a representation of the user in inference from a partial history, predictions and evaluation are performed on the \textit{fold-out} set. In our experiments, the fold-in is composed of 80\% of the randomly sampled interactions of a user, the rest is for the fold-out. This data split is illustrated on the right side of~\cref{fig:figure_protocol}.

\subsection{Baselines and implementation details}
\subsubsection{Baselines}
We use several GNN architectures GCN (\cite{Kipf2016Semi-SupervisedNetworks}), GAT (\cite{Velickovic2017GraphNetworks}), GIN (\cite{Xu2018HowNetworks}) and LightGCN (\cite{He2020LightGCN:Recommendation}) and compare their original training, \ie with the BPR (\cite{Rendle2012BPR:Feedback}), to \textsc{ITEM} training. To assess our model against the state of the art, we compare it with various CF model families, such as MF (MF-BPR (\cite{Rendle2012BPR:Feedback}), ENMF (\cite{Chen2020EfficientRecommendation})), metric-learning based (CML (\cite{Hsieh2017CollaborativeLearning})), graph embedding (DeepWalk (\cite{Perozzi2014DeepWalk:Representations}), LINE (\cite{Tang2015LINE:Embedding}), Node2Vec (\cite{Grover2016Node2vec:Networks})), VAE models (Mult-VAE (\cite{Liang2018VariationalFiltering})), MP-GNNs (NGCF (\cite{Wang2019NeuralFiltering}), DGCF (\cite{Wang2020DisentangledFiltering}), LightGCN (\cite{He2020LightGCN:Recommendation}), NIA-GCN (\cite{Sun2020NeighborRecommendation}), LR-GCCF (\cite{Chen2020RevisitingApproach})). \textcolor{black}{We also compare ITEM to a recent embedding enhancement method \cite{Chen2023Adap-tau:Recommendation}.}\\ 
 Finally, we compare \textsc{ITEM} to other losses and negative sampling methods such as NeuralNDCG (\cite{https://doi.org/10.48550/arxiv.2102.07831}, another ranking loss); SGL-ED (\cite{Wu2021Self-supervisedRecommendation}) which combines \textsc{BPR} and a self-supervised loss; XSimGCL and SimGCL (\cite{Yu2022XSimGCL:Recommendation}) two contrastive-based models for CF;  MixGCF (\cite{Huang2021MixGCF:Systems}) which employs a hard negative sampling method for \textsc{BPR}. For a fair comparison of the various loss functions, we employ the same LightGCN backbone across all models.
 
\subsubsection{Implementation details}
We give all details on hyper-parameters and optimization procedure in Sec. B.2 of supplementary. For the inductive setting, we found experimentally that learning user embeddings during training is harmful for the generalization of the GNNs to new users, so for training and evaluation users' embeddings are inferred using message passing only.

\subsection{Main results}

In this section we present our main results, we compare \textsc{ITEM} vs state-of-the-art methods. We first compare \textsc{ITEM} on the standard transductive protocol in~\cref{tab:sota_transductive}, we then compare \textsc{ITEM} on the indcutive protocol in~\cref{tab:sota_inductive}. For both protocols we use a LightGCN (\cite{He2020LightGCN:Recommendation}) backbone following (\cite{Huang2021MixGCF:Systems,Wu2021Self-supervisedRecommendation}).

\subsubsection{Transductive state of the art comparison} 

In~\cref{tab:sota_transductive} we compare, in the transductive setting, \textsc{ITEM} using a LightGCN (\cite{He2020LightGCN:Recommendation}) backbone vs state-of-the-art methods.
We show that across all datasets \textsc{ITEM} outperforms all the MP-GNNs state-of-the-art methods NGCF, LR-GCCF, NIA-GCCN, LightGCN and DGCF. It surpasses non-graph methods on the three datasets with relative improvements of +18\% \textsc{NDCG}@20 on Yelp-2018 or 21\% \textsc{NDCG}@20 on Amazon-book vs Mult-VAE, which shows the interest of using dedicated architectures for graph learning. Furthermore, \textsc{ITEM} outperforms the recent NeuralNDCG (\cite{https://doi.org/10.48550/arxiv.2102.07831}), \eg +0.89 R@20 and +0.65 \textsc{NDCG}@20 on Yelp-2018, as well as SGL-ED, \eg +0.55 R@20 and +0.23 \textsc{NDCG}@20 on Amazon-Book, and the advanced sampling method MixGCF (\cite{Huang2021MixGCF:Systems}), \eg +2.45 R@20 and +2.64 \textsc{NDCG}@20 on MovieLens-1M. Despite the limitations of MP-GNNs (\cite{Alon2020OnImplications,Chen2019MeasuringView}), \textsc{ITEM} shows competitive results against the hybrid model SimpleX (\cite{Mao2021SimpleX:Filtering}) with better results on two of three datasets with +1.78 R@20 on MovieLens-1M and +0.38 R@20 on Yelp-2018 with the best state of the arts results.

\begin{table*}[ht]
    \setlength\tabcolsep{0.3pt}
    \caption{Comparison of \textsc{ITEM} vs state-of-the-art methods on three \textit{\textbf{inductive}} benchmarks. Best results in \textbf{bold}.}
    \label{tab:sota_inductive} 
    \centering
    \begin{tabularx}{\textwidth}{l YY | YY | YY }
        \toprule
         \multirow{2}{*}{Method} & \multicolumn{2}{c|}{MovieLens-100K} & \multicolumn{2}{c|}{Yelp-2018} & \multicolumn{2}{c}{Amazon-Book} \\
        \cmidrule{2-7}
         &  R@20 &  N@20 & R@20 &  N@20 & R@20 & N@20 \\
         \midrule
         Mult-VAE (\cite{Liang2018VariationalFiltering})& 30.14 & 28.28  &  10.15 &  8.18 & 10.86 & 9.2  \\

         \midrule
         
         GCN (\cite{Kipf2016Semi-SupervisedNetworks})&28.74 &27.68& 7.34 & 5.76 & 8.85 & 7.61 \\
         GAT (\cite{Velickovic2017GraphNetworks})& 31.01 & 28.92  & 9.04 & 7.32 & 9.88 & 8.17  \\
         GIN (\cite{Xu2018HowNetworks})&29.71 & 27.58  & 7.34 & 5.76 & 9.62  & 8.05 \\
         LightGCN (\cite{He2020LightGCN:Recommendation}) & 30.79 &  29.73 & 7.88 & 6.34 & 9.56 & 8.02\\
         \midrule
         NeuralNDCG (\cite{https://doi.org/10.48550/arxiv.2102.07831})&31.12 &30.07 & 9.14 & 7.62 & 9.54 & 8.19 \\
         
         MixGCF (\cite{Huang2021MixGCF:Systems})& 32.07 &30.62 & 9.85 & 8.21 & 10.11 & 9.63 \\
         
         \midrule
         \rowcolor[gray]{.95} \textbf{ITEM (ours)} & \textbf{33.84} & \textbf{32.63} & \textbf{10.54} & \textbf{8.70} & \textbf{11.03} & \textbf{9.89}  \\
         \bottomrule
    \end{tabularx}
\end{table*}

\begin{table*}[!h]
    \setlength\tabcolsep{0.3pt}
    \caption{Comparison of \textsc{ITEM} vs state-of-the-art methods on three \textit{\textbf{transductive}} benchmarks. Best results in \textbf{bold}.}
    \label{tab:sota_transductive} 
    \centering
    \begin{tabularx}{\textwidth}{l YY | YY | YY }
        \toprule
         \multirow{2}{*}{Method} & \multicolumn{2}{c|}{MovieLens-1M} & \multicolumn{2}{c|}{Yelp-2018} & \multicolumn{2}{c}{Amazon-Book} \\
        \cmidrule{2-7}
         &  R@20 &  N@20 & R@20 &  N@20 & R@20 & N@20 \\
        \midrule
        MF-BPR (\cite{Rendle2012BPR:Feedback}) & 21.53 & 21.75 & 5.49 & 4.45 & 3.38 & 2.61 \\
        CML (\cite{Hsieh2017CollaborativeLearning})& 17.30 & 15.63 & 6.22 & 5.36 & 5.22 & 4.28 \\
        ENMF (\cite{Chen2020EfficientRecommendation})  & 23.15 & 20.69 & 6.24 & 5.15 & 3.59 & 2.81 \\
         
         \midrule
         DeepWalk (\cite{Perozzi2014DeepWalk:Representations}) & 13.48 & 10.57 & 4.76& 3.78 & 3.46 & 2.64 \\
         LINE (\cite{Tang2015LINE:Embedding})  & 23.36 & 22.26 & 5.49 & 4.46 & 4.10 & 3.18 \\
         Node2Vec (\cite{Grover2016Node2vec:Networks}) & 14.75 & 11.86 & 4.52 & 3.60 & 4.02 & 3.09 \\
    
         \midrule
         Mult-VAE (\cite{Liang2018VariationalFiltering})& 29.23 & 23.84 &  6.41 &  4.97 & 4.46 & 3.33 \\
         
         \midrule
         NGCF (\cite{Wang2019NeuralFiltering}) & 25.13 & 25.11 & 5.79 & 4.77 & 3.44 & 2.63 \\
         LR-GCCF (\cite{Chen2020RevisitingApproach})  & 22.31&21.24&5.61&3.43&3.35&2.65 \\
         NIA-GCN (\cite{Sun2020NeighborRecommendation}) & 23.59&22.42&5.99&4.91&3.69&2.87\\
         LightGCN (\cite{He2020LightGCN:Recommendation}) & 25.76 &  24.27 & 6.28 & 5.15 & 4.23 & 3.17 \\ 
         DGCF (\cite{Wang2020DisentangledFiltering}) & 26.40 & 25.04 & 6.54 & 5.34 & 4.22 & 3.24 \\
         
         \midrule
         
         SGL-ED (\cite{Wu2021Self-supervisedRecommendation})& 26.34 & 24.87  & 6.75 & 5.55 & 4.78 & 3.79  \\
         SimGCL (\cite{Yu2022XSimGCL:Recommendation})& 27.55 & 25.01 & 7.21 & 6.01 &4.90 & 3.98\\ 
         XSimGCL (\cite{Yu2022XSimGCL:Recommendation}) & 27.94 & 25.36 & 7.33 & 6.06 &5.02 &4.11 \\
         Adap-$\tau$ \cite{Chen2023Adap-tau:Recommendation}&27.87 &26.15 & 7.33 &\textbf{ 6.12} &\textbf{ 6.12} & \textbf{4.90} \\
         NeuralNDCG (\cite{https://doi.org/10.48550/arxiv.2102.07831})& 29.45 &25.13 & 6.50 & 5.23 & 4.38 & 3.27 \\
         
         MixGCF (\cite{Huang2021MixGCF:Systems})& 27.35 & 24.56 & 7.17 & 5.84 & 4.51 & 3.41 \\
         \midrule
         \rowcolor[gray]{.95} \textbf{ITEM (ours)} & \textbf{29.80} & \textbf{27.20} & \textbf{7.39} & {5.88} & {5.23} & {4.02}  \\
         \bottomrule
    \end{tabularx}
\end{table*}

\subsubsection{Inductive state-of-the-art comparison:}
MP-GNNs can be easily evaluated in an inductive setup, unlike matrix factorization models or random walk embedding methods. MP-GNNs have a strong ability to generalize to users not seen during the training phase with the propagation and update process. We show that our \textsc{ITEM} method significantly boosts the performance of MP-GNNs in the inductive setup. We compare \textsc{ITEM} using LightGCN (\cite{He2020LightGCN:Recommendation}) to state-of-the-art methods, and show that on the three datasets, \textsc{ITEM} sets new state-of-the-art performances for inductive recommendation. It outperforms Mult-VAE (\cite{Liang2018VariationalFiltering}) which is designed for the inductive setting, by {+3.79 \textsc{NDCG}@20} on MovieLens-100k, +1.4 \textsc{NDCG}@20 on Yelp-2018 and +0.69 \textsc{NDCG}@20 on Amazon-book.  We compare \textsc{ITEM} to different GNNs that were optimized using the \textsc{BPR} loss, this is further studied in Tab. 1 of supplementary where we show the interest of the well designed loss of \textsc{ITEM} vs the standard \textsc{BPR} loss. Finally \textsc{ITEM} outperforms MixGCF (\cite{Huang2021MixGCF:Systems}) on the three datasets, \eg on MovieLens-100k with +1.77 R@20 or +1.58 R@20 on Yelp-2018 as well as the ranking loss function NeuralNDCG (\cite{https://doi.org/10.48550/arxiv.2102.07831}), \eg +1.54 R@20 and +1.08 \textsc{NDCG}@20. Although message-passing models face certain challenges (\cite{Alon2020OnImplications,Chen2019MeasuringView}), when evaluated with the proposed inductive protocol they outperform state-of-the-art methods and exhibit fast training and strong generalization capabilities for unseen users during the training phase.

\subsection{Ablation studies}\label{sec:exp_transductive_learning}

In this section we perform ablation studies for the two elements of \textsc{ITEM}. We first study the impact of using $\llargo$ vs the \textsc{BPR} loss in~\cref{tab:main_transductive}. We then study the impact of the sampling in~\cref{tab:sampling_compa}.

\subsubsection{Ranking loss vs \textsc{BPR}} In~\cref{tab:main_transductive} we compare in the same settings the \textsc{BPR} loss (\cite{Rendle2012BPR:Feedback}) and our proposed ranking-based loss $\llargo$ on three \textit{transductive} benchmarks, and show that for all four considered architectures, \textsc{ITEM} outperforms the \textsc{BPR} loss. Using our loss in combination with LightGCN (\cite{He2020LightGCN:Recommendation}), the state-of-the-art GNN for transductive datasets, increases relative performances from +10.05\% \textsc{NDCG}@20 on MovieLens up to +22.1\% \textsc{NDCG}@20 on Amazon-book. 
We can also note that when using \textsc{ITEM} more expressive GNNs, \eg GAT (\cite{Velickovic2017GraphNetworks}), perform better than LightGCN on MovieLens-1M and Amazon-book. With GAT the relative improvements range from +18.55\% \textsc{NDCG}@20 on MovieLens-1M up to +54.9\% \textsc{NDCG}@20 on Amazon-book. Overall, \cref{tab:main_transductive} shows the huge benefits of optimizing a ranking-based loss rather than a \textit{proxy} loss. Table 1 of supplementary shows similar results for the inductive benchmarks. 
In Table 1 of supplementary we show that the best performing architectures on the inductive setting are not the same as in the transductive setting.

\begin{table*}[ht]
    \setlength\tabcolsep{1pt}
    \setlength\extrarowheight{4pt}
    \caption{Comparison of our ranking-based loss \textsc{ITEM} vs the \textsc{BPR} loss (\cite{Rendle2012BPR:Feedback}), using different GNN architectures on 3 \textit{\textbf{transductive}} benchmarks. For each architecture best results is \textbf{bold}, best overall results is \underline{underlined}. }
    \label{tab:main_transductive} 
    \centering
    \begin{tabularx}{\textwidth}{l l *{2}{Y|Y|} Y|Y}
        \toprule
         & \multirow{2}{*}{Method} & \multicolumn{2}{c|}{MovieLens-1M} & \multicolumn{2}{c|}{Yelp-2018} & \multicolumn{2}{c}{Amazon-book} \\
        \cmidrule{3-8}
         && R@20 & NDCG@20 & R@20 & NDCG@20 & R@20 & NDCG@20 \\
         \midrule
         \multirow{3}{*}{\rotatebox[origin=c]{90}{GCN}}
         & BPR & 27.26 & 22.75 & 5.21 & 4.04 & 3.48 & 2.55 \\
         & $\llargo$  &\textbf{29.96} &\textbf{25.62}& \textbf{6.81} & \textbf{5.33} & \textbf{4.89} & \textbf{3.70} \\
         & \small \%Improv. & \small {+9.90\%} & \small {+12.62\%} & \small  {+30.7\%} & \small  {+31.9\%} & \small  {+40.5\%} & \small  {+45.1\%} \\
         \midrule
         \multirow{3}{*}{\rotatebox[origin=c]{90}{GIN}}
         & BPR & 26.99 &22.28  & 5.52 & 4.32 & 3.42 & 2.54 \\
          & $\llargo$ &\textbf{29.94}& \textbf{25.60} & \textbf{6.87} & \textbf{5.38} & \underline{\textbf{5.24}} & \underline{\textbf{3.94}}  \\
         &  \small \%Improv. & \small {+10.93\%} & \small {+14.90\%} & \small  {+24.5\%} & \small  {+24.5\%} & \small  {+53.2\%} & \small  {+55.1\%} \\
        \midrule
         \multirow{3}{*}{\rotatebox[origin=c]{90}{GAT}}
         & BPR &  27.50& 22.96 & 5.08 & 3.89 & 3.41 & 2.53\\
         &  $\llargo$ & \underline{\textbf{30.43}} & \underline{\textbf{27.22}} & \textbf{7.03} & \textbf{5.47} & \underline{\textbf{5.20}} & \underline{\textbf{3.92}} \\
        &  \small \%Improv. & \small {+10.65\%} & \small {+18.55\%} & \small {+38.4\%} & \small  {+40.6\%} & \small  {+52.5\%} & \small  {+54.9\%} \\
         \midrule
         \multirow{3}{*}{\rotatebox[origin=c]{90}{LGCN~}}
         & BPR & 25.76 & 24.27 & 6.26 & 5.14 & 4.23 & 3.17 \\
          &  $\llargo$ & \textbf{29.69} & \textbf{26.71} & \underline{\textbf{7.25}} & \underline{\textbf{5.70}} & \textbf{5.09} & \textbf{3.87} \\
          &  \small \%Improv. & \small {+15.27\%} & \small{+10.05\%} & \small{+15.81\%} & \small{+10.89\%} & \small{+20.3\%} & \small{+22.1\%}\\
         \bottomrule
    \end{tabularx}
\end{table*}

\subsubsection{Negative sampling} We show in~\cref{tab:sampling_compa} ablation studies for the impact of our PPR sampling strategy. On both datasets our sampling strategy boosts the performances of \textsc{ITEM}, \eg +3.2\% \textsc{NDCG}@20 relative improvement on Yelp-2018 and +3.9\% \textsc{NDCG}@20 on Amazon-book. We also use the MixGCF (\cite{Huang2021MixGCF:Systems}) negative sampling with \textsc{ITEM}, however due to its high computation overhead, we cannot sample many negatives. This leads to MixGCF negatively affecting performances on Yelp-2018 and Amazon-book, \eg~-0.34 \textsc{NDCG}@20 on Amazon-book.

\begin{table*}[ht]
    \caption{
    Ablation study of the components of \textsc{ITEM}. The loss: \textcolor{black}{\cmark} for $\llargo$, \textsc{BPR} otherwise (\xmark). And sampling: random negative sampling vs MixGCF (\cite{Huang2021MixGCF:Systems}) vs the \textsc{ITEM} sampling~\cref{eq:pprneg} (PPR).With LightGCN (\cite{He2020LightGCN:Recommendation}) on \textit{\textbf{transductive}} benchmarks.
    }
    \label{tab:sampling_compa} 
    \centering
    \begin{tabularx}{\textwidth}{ccc YY | YY }
        \toprule
         \multirow{2}{*}{$\llargo$} & \multirow{2}{*}{MixGCF} & \multirow{2}{*}{PPR (ours)} & \multicolumn{2}{c|}{Yelp-2018} & \multicolumn{2}{c}{Amazon-book} \\
         \cmidrule{4-7}
         &&& R@20 & NDCG@20 & R@20 & NDCG@20 \\
         \midrule
         \xmark &\xmark &\xmark & 6.26 & 5.14 & 4.23 & 3.17 \\
         \textcolor{black}{\cmark} & \xmark & \xmark & 7.25  & 5.70 & 5.09 & 3.87 \\
         \textcolor{black}{\cmark} & \textcolor{black}{\cmark} & \xmark & 7.21 & 5.71 & 4.86 & 3.53\\
         \rowcolor[gray]{.95} \textcolor{black}{\cmark} & \xmark & \textcolor{black}{\cmark} &  \textbf{7.39} & \textbf{5.88} & \textbf{5.23}  & \textbf{4.02}   \\
         \bottomrule
    \end{tabularx}
\end{table*}


\subsection{Model analysis}\label{sec:model_analysis}
\subsubsection{Efficiency comparison}
\begin{wraptable}{r}{0.47\textwidth}
    \vspace{-\intextsep}
    \caption{Computational cost on Yelp-2018 with LightGCN backbone.}
    \label{tab:time_compa} 
    \setlength\tabcolsep{0pt}
    \begin{tabularx}{0.47\textwidth}{ l | YY}
        \toprule
         Method & Time / epoch & Training Time \\
         \midrule
         BPR & 53s & 10h26 \\
         MixGCF & 145s  & 6h07 \\
         ITEM (ours) & \textbf{12.3s} & \textbf{1h43} \\
         \bottomrule
    \end{tabularx}
\end{wraptable}
In \cref{tab:time_compa} we compare the training times of LightGCN (\cite{He2020LightGCN:Recommendation}) using the \textsc{BPR} loss, MixGCF and \textsc{ITEM} on the Yelp-2018 dataset until convergence.  We train these 3 variants with a Quadro RTX 5000 GPU. LightGCN-\textsc{BPR} takes 10h26 before convergence on the Yelp-2018 dataset, which contains 1.5 million interactions. With MixGCF, the convergence time is reduced to approximately 6 hours, as hard negative sampling accelerates convergence, despite the time per epoch being more than twice as long due to the online generation of the hard negatives examples. Finally, \textsc{ITEM} leads to model convergence in just 1 hour and 43 minutes. \textsc{ITEM} performs far fewer iterations per epoch by building batches of negative and positive items per user to approximate the \textsc{NDCG} metric.
\subsubsection{Impact of $\tau$}
\begin{figure}[ht]
    \centering
        
    \begin{subfigure}[t]{0.45\textwidth}
        \includegraphics[width=\textwidth]{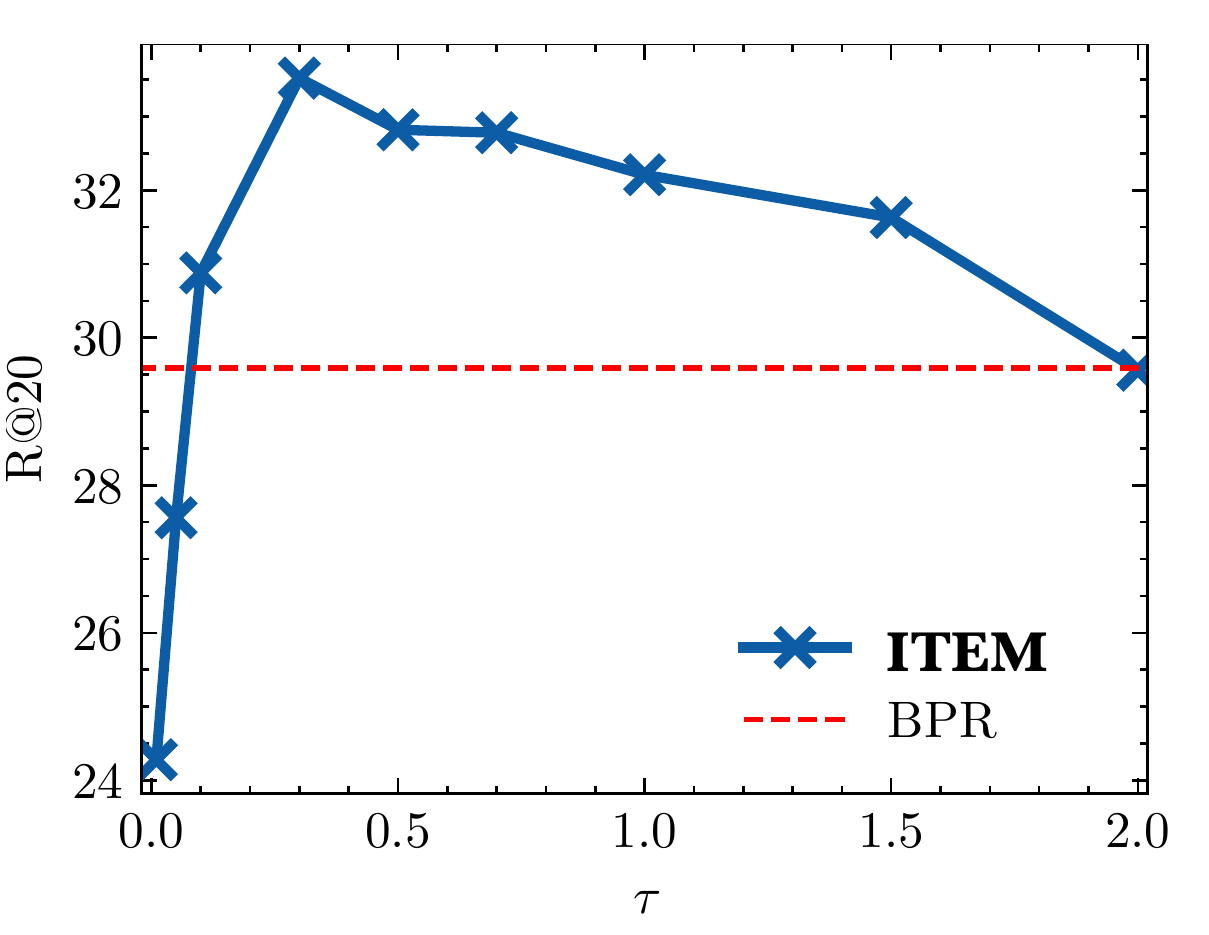}
        \caption{$R@20$ vs $\tau$ in~\cref{eq:def_approx_rank}.
        }
        \label{fig:tau_recall}
    \end{subfigure}
    ~
    \begin{subfigure}[t]{0.45\textwidth}
    \includegraphics[width=\textwidth]{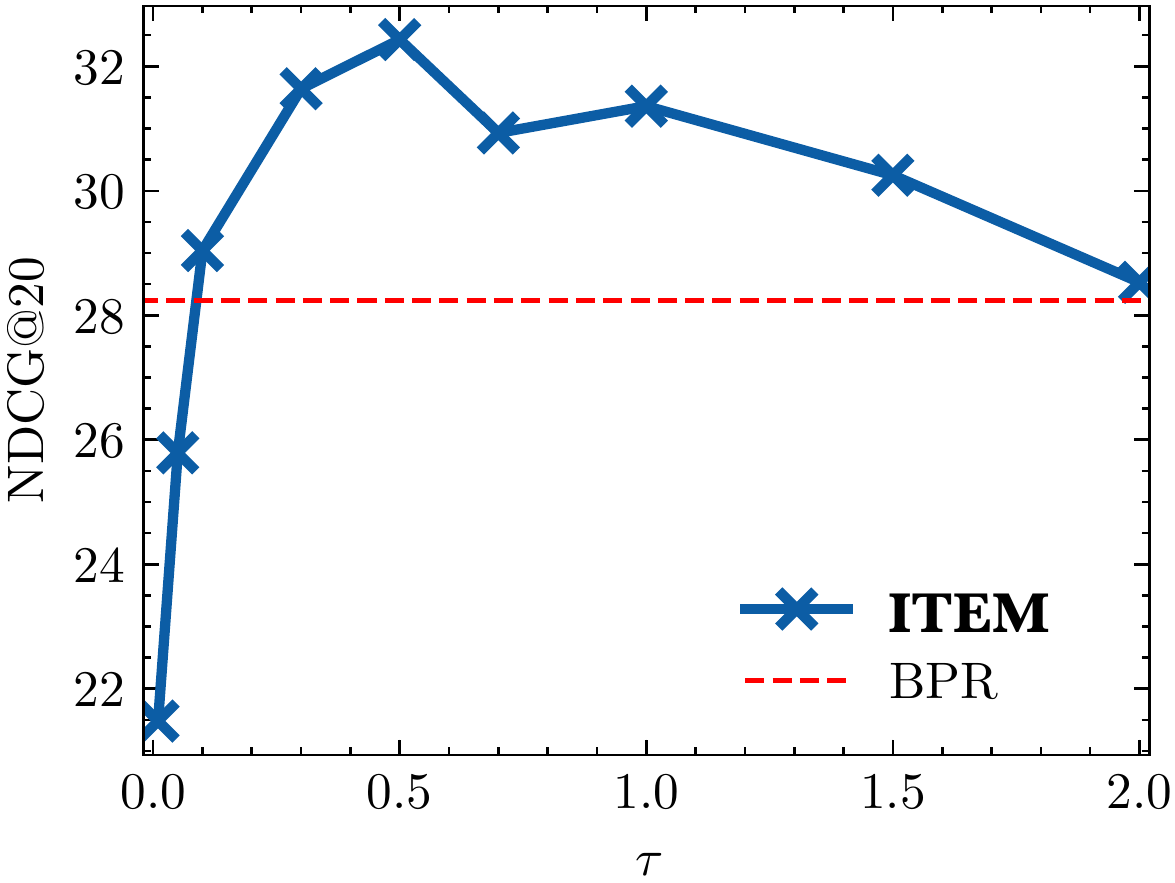}
    \caption{$\NDCG@20$ vs $\tau$ in~\cref{eq:def_approx_rank}.
    }
    \label{fig:tau_ndcg}
    \end{subfigure}%

    \caption{
    $\tau$ in~\cref{eq:def_approx_rank} vs R@20, \textsc{NDCG}@20 on MovieLens-100k (inductive) with LightGCN (\cite{He2020LightGCN:Recommendation}).
    }
    \label{fig:ablation_tau}
\end{figure}

We show in~\cref{fig:tau_recall,fig:tau_ndcg} the impact of $\tau$ in~\cref{eq:def_approx_rank} on the R@20 and \textsc{NDCG}@20 when optimizing \textsc{ITEM} on MovieLens for the inductive protocol. We can see on both figures that our method is robust for a wide range of $\tau$. Specifically, for values of $\tau$ in range 0.1 to 2.0, \textsc{ITEM} outperforms the \textsc{BPR} loss. Also note that using a finer selection of $\tau$ could lead to better results than reported in~\cref{tab:sota_inductive}, \eg using $\tau=0.3$ leads to a R@20 of 33.52 in~\cref{fig:tau_recall} against 33.13 for the value of $\tau$ used in~\cref{tab:sota_inductive} (note that in~\cref{tab:sota_inductive} \textsc{ITEM} additionally uses the PPR sampling).

\subsection{Qualitative results}

\begin{figure}[ht]
     \centering
    \includegraphics[width=0.8\textwidth]{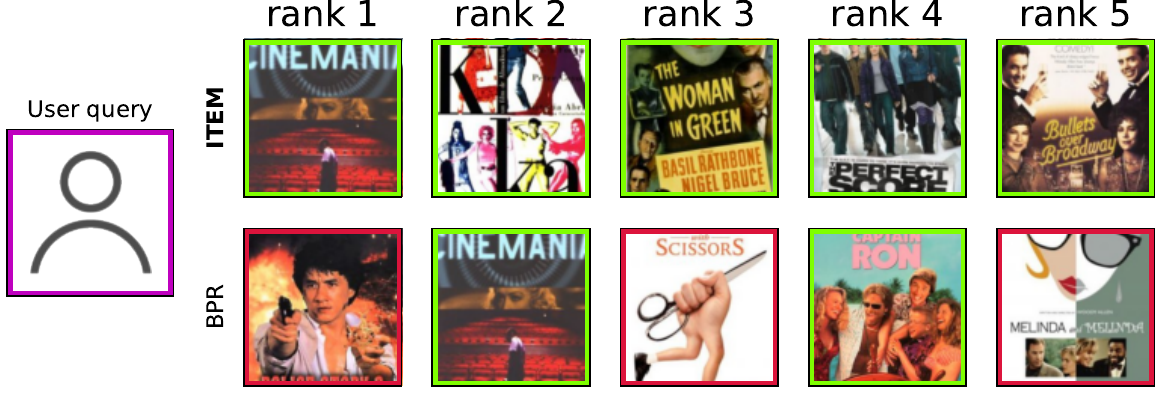}
     \caption{Qualitative results on MovieLens-100k. We compare the ranking obtained using LightGCN trained with the baseline \textsc{BPR} loss (bottom row) to the ranking obtained using \textbf{\textsc{ITEM}} (top row). Positive elements are highlighted in green, while negative elements are indicated in red.}
     \label{fig:qual_results}
\end{figure}

In the qualitative results presented in~\cref{fig:qual_results}, a comparative analysis between our approach and the \textsc{BPR} loss baseline reveals the superior ranking achieved by \textsc{ITEM}. This enhanced ranking is not solely attributable to optimizing the target metric directly during training; it is also a result of the efficacy of our hard negative sampling strategy. This strategy contributes to a more effective differentiation between negative and positive instances, further refining the ranking quality. Supplementary B.6 includes a detailed listing of the top-20 results for comprehensive examination.

\section{Conclusion} \label{sec:conclusion}
In this study, we introduce the ITEM framework to optimize GNNs for the top-$k$ recommendation task. By incorporating ranking losses used in image retrieval and directly optimizing item ranking for a given user, we demonstrate the limitations of the BPR loss, a widely-used pairwise loss within the GNN community. 

Our proposed list-wise loss requires sampling a substantial number of hard {\em negative} items for each user, which is a challenging task. Unable to employ {\em online} methods such as MixGCF (\cite{Huang2021MixGCF:Systems}), we devised an {\em offline} strategy to sample negatives, relying on the Personalized PageRank score. It leverages the graph structure of a local neighborhood around a user: an item with a high PPR score for a user will be more challenging to distinguish for an MP-GNN, and enhance the learning performance.

Furthermore, we expanded the evaluation of GNNs by benchmarking on an inductive user-split protocol, which gauges the generalization capacity of GNNs. In both inductive and transductive protocols, we showcased the effectiveness of our ITEM learning framework, exhibiting improvements over the conventional BPR loss function, even compared to   standard GNN training-enhancements methods such as SGL-ED, SIMGCL, or MixGCF. Our findings demonstrate that using ranking losses to  enhance the training of MP-GNNs is very a promising research direction.

\bibliography{main}

\begin{thebibliography}{50}
\providecommand{\natexlab}[1]{#1}
\providecommand{\url}[1]{\texttt{#1}}
\expandafter\ifx\csname urlstyle\endcsname\relax
  \providecommand{\doi}[1]{doi: #1}\else
  \providecommand{\doi}{doi: \begingroup \urlstyle{rm}\Url}\fi

\bibitem[Alon \& Yahav(2020)Alon and Yahav]{Alon2020OnImplications}
Uri Alon and Eran Yahav.
\newblock {On the Bottleneck of Graph Neural Networks and its Practical Implications}.
\newblock 2020.
\newblock \doi{10.48550/arxiv.2006.05205}.

\bibitem[Asghar(2016)]{asghar2016yelp}
Nabiha Asghar.
\newblock Yelp dataset challenge: Review rating prediction.
\newblock 2016.

\bibitem[Brown et~al.(2020)Brown, Xie, Kalogeiton, and Zisserman]{brown2020smooth}
Andrew Brown, Weidi Xie, Vicky Kalogeiton, and Andrew Zisserman.
\newblock {Smooth-AP: Smoothing the Path Towards Large-Scale Image Retrieval}.
\newblock 2020.

\bibitem[Bruch et~al.(2019)Bruch, Zoghi, Bendersky, and Najork]{bruch2019revisiting}
Sebastian Bruch, Masrour Zoghi, Michael Bendersky, and Marc Najork.
\newblock Revisiting approximate metric optimization in the age of deep neural networks.
\newblock In \emph{ACM SIGIR}, 2019.

\bibitem[Burges et~al.(2005)Burges, Shaked, Renshaw, Lazier, Deeds, Hamilton, and Hullender]{ranknet}
Chris Burges, Tal Shaked, Erin Renshaw, Ari Lazier, Matt Deeds, Nicole Hamilton, and Greg Hullender.
\newblock Learning to rank using gradient descent.
\newblock In \emph{{ACM ICML}}, 2005.

\bibitem[Burges et~al.(2006)Burges, Ragno, and Le]{lambdarank}
Christopher Burges, Robert Ragno, and Quoc Le.
\newblock Learning to rank with nonsmooth cost functions.
\newblock In \emph{NeurIPS}, 2006.

\bibitem[Cao et~al.(2007)Cao, Qin, Liu, Tsai, and Li]{10.1145/1273496.1273513}
Zhe Cao, Tao Qin, Tie-Yan Liu, Ming-Feng Tsai, and Hang Li.
\newblock Learning to rank: From pairwise approach to listwise approach.
\newblock In \emph{{ACM ICML}}, 2007.

\bibitem[Chen et~al.(2020{\natexlab{a}})Chen, Zhang, Zhang, Liu, and Ma]{Chen2020EfficientRecommendation}
Chong Chen, Min Zhang, Yongfeng Zhang, Yiqun Liu, and Shaoping Ma.
\newblock {Efficient neural matrix factorization without sampling for recommendation}.
\newblock \emph{ACM Transactions on Information Systems}, 38\penalty0 (2):\penalty0 14, 1 2020{\natexlab{a}}.
\newblock ISSN 15582868.
\newblock \doi{10.1145/3373807}.
\newblock URL \url{https://doi.org/10.1145/3373807}.

\bibitem[Chen et~al.(2019)Chen, Lin, Li, Li, Zhou, and Sun]{Chen2019MeasuringView}
Deli Chen, Yankai Lin, Wei Li, Peng Li, Jie Zhou, and Xu~Sun.
\newblock {Measuring and Relieving the Over-smoothing Problem for Graph Neural Networks from the Topological View}.
\newblock \emph{AAAI 2020 - 34th AAAI Conference on Artificial Intelligence}, pp.\  3438--3445, 9 2019.
\newblock ISSN 2159-5399.
\newblock \doi{10.1609/aaai.v34i04.5747}.
\newblock URL \url{https://arxiv.org/abs/1909.03211v2}.

\bibitem[Chen et~al.(2023)Chen, Wu, Wu, Zhou, Cao, and He]{Chen2023Adap-tau:Recommendation}
Jiawei Chen, Junkang Wu, Jiancan Wu, Sheng Zhou, Xuezhi Cao, and Xiangnan He.
\newblock {Adap-{\$}{\textbackslash}tau{\$}: Adaptively Modulating Embedding Magnitude for Recommendation}.
\newblock \emph{ACM Web Conference 2023 - Proceedings of the World Wide Web Conference, WWW 2023}, 1:\penalty0 1085--1096, 2 2023.
\newblock \doi{10.1145/3543507.3583363}.
\newblock URL \url{http://arxiv.org/abs/2302.04775 http://dx.doi.org/10.1145/3543507.3583363}.

\bibitem[Chen et~al.(2020{\natexlab{b}})Chen, Wu, Hong, Zhang, and Wang]{Chen2020RevisitingApproach}
Lei Chen, Le~Wu, Richang Hong, Kun Zhang, and Meng Wang.
\newblock {Revisiting Graph based Collaborative Filtering: A Linear Residual Graph Convolutional Network Approach}.
\newblock \emph{AAAI 2020 - 34th AAAI Conference on Artificial Intelligence}, pp.\  27--34, 1 2020{\natexlab{b}}.
\newblock ISSN 2159-5399.
\newblock \doi{10.1609/aaai.v34i01.5330}.
\newblock URL \url{https://arxiv.org/abs/2001.10167v1}.

\bibitem[Croft et~al.(2010)Croft, Metzler, and Strohman]{croft2010search}
W~Bruce Croft, Donald Metzler, and Trevor Strohman.
\newblock \emph{Search engines: Information retrieval in practice}, volume 520.
\newblock Addison-Wesley Reading, 2010.

\bibitem[Fey \& Lenssen(2019)Fey and Lenssen]{Fey/Lenssen/2019}
Matthias Fey and Jan~E. Lenssen.
\newblock Fast graph representation learning with {PyTorch Geometric}.
\newblock In \emph{ICLR Workshop on Representation Learning on Graphs and Manifolds}, 2019.

\bibitem[Gilmer et~al.(2017)Gilmer, Schoenholz, Riley, Vinyals, and Dahl]{Gilmer2017NeuralChemistry}
Justin Gilmer, Samuel~S. Schoenholz, Patrick~F. Riley, Oriol Vinyals, and George~E. Dahl.
\newblock {Neural Message Passing for Quantum Chemistry}.
\newblock 4 2017.
\newblock URL \url{http://arxiv.org/abs/1704.01212}.

\bibitem[Grover \& Leskovec(2016)Grover and Leskovec]{Grover2016Node2vec:Networks}
Aditya Grover and Jure Leskovec.
\newblock {node2vec: Scalable Feature Learning for Networks}.
\newblock 7 2016.
\newblock URL \url{https://arxiv.org/abs/1607.00653}.

\bibitem[Hadsell et~al.(2006)Hadsell, Chopra, and LeCun]{hadsell2006dimensionality}
Raia Hadsell, Sumit Chopra, and Yann LeCun.
\newblock Dimensionality reduction by learning an invariant mapping.
\newblock In \emph{{IEEE} CVPR}, 2006.

\bibitem[Hamilton et~al.(2017)Hamilton, Ying, and Leskovec]{Hamilton2017InductiveGraphs}
William~L. Hamilton, Rex Ying, and Jure Leskovec.
\newblock {Inductive Representation Learning on Large Graphs}.
\newblock 6 2017.
\newblock URL \url{https://arxiv.org/abs/1706.02216}.

\bibitem[Harper \& Konstan(2015)Harper and Konstan]{harper2015movielens}
F~Maxwell Harper and Joseph~A Konstan.
\newblock The {Movielens} datasets: History and context.
\newblock \emph{{ACM TIIS}}, 2015.

\bibitem[He et~al.(2017)He, Liao, Zhang, Nie, Hu, and Chua]{He2017NeuralFiltering}
Xiangnan He, Lizi Liao, Hanwang Zhang, Liqiang Nie, Xia Hu, and Tat-Seng Chua.
\newblock {Neural Collaborative Filtering}.
\newblock 8 2017.
\newblock URL \url{http://arxiv.org/abs/1708.05031}.

\bibitem[He et~al.(2020)He, Deng, Wang, Li, Zhang, and Wang]{He2020LightGCN:Recommendation}
Xiangnan He, Kuan Deng, Xiang Wang, Yan Li, Yongdong Zhang, and Meng Wang.
\newblock {LightGCN: Simplifying and Powering Graph Convolution Network for Recommendation}.
\newblock 2 2020.
\newblock URL \url{https://arxiv.org/abs/2002.02126}.

\bibitem[Hsieh et~al.(2017)Hsieh, Yang, Cui, Lin, Belongie, and Estrin]{Hsieh2017CollaborativeLearning}
Cheng~Kang Hsieh, Longqi Yang, Yin Cui, Tsung~Yi Lin, Serge Belongie, and Deborah Estrin.
\newblock {Collaborative metric learning}.
\newblock \emph{26th International World Wide Web Conference, WWW 2017}, pp.\  193--201, 2017.
\newblock \doi{10.1145/3038912.3052639}.

\bibitem[Huang et~al.(2021)Huang, Dong, Ding, Yang, Feng, Wang, and Tang]{Huang2021MixGCF:Systems}
Tinglin Huang, Yuxiao Dong, Ming Ding, Zhen Yang, Wenzheng Feng, Xinyu Wang, and Jie Tang.
\newblock {MixGCF: An Improved Training Method for Graph Neural Network-based Recommender Systems}.
\newblock \emph{Proceedings of the ACM SIGKDD International Conference on Knowledge Discovery and Data Mining}, pp.\  665--674, 8 2021.
\newblock \doi{10.1145/3447548.3467408}.

\bibitem[J{\"a}rvelin \& Kek{\"a}l{\"a}inen(2002)J{\"a}rvelin and Kek{\"a}l{\"a}inen]{jarvelin2002cumulated}
Kalervo J{\"a}rvelin and Jaana Kek{\"a}l{\"a}inen.
\newblock Cumulated gain-based evaluation of {IR} techniques.
\newblock \emph{ACM TOIS}, 2002.

\bibitem[Kipf \& Welling(2016)Kipf and Welling]{Kipf2016Semi-SupervisedNetworks}
Thomas~N. Kipf and Max Welling.
\newblock {Semi-Supervised Classification with Graph Convolutional Networks}.
\newblock 9 2016.
\newblock URL \url{http://arxiv.org/abs/1609.02907}.

\bibitem[Li et~al.(2017)Li, Wang, Zhang, Li, and Zuo]{li2017joint}
Mu~Li, Qilong Wang, David Zhang, Peihua Li, and Wangmeng Zuo.
\newblock Joint distance and similarity measure learning based on triplet-based constraints.
\newblock \emph{IS}, 2017.

\bibitem[Liang et~al.(2018)Liang, Krishnan, Hoffman, and Jebara]{Liang2018VariationalFiltering}
Dawen Liang, Rahul~G. Krishnan, Matthew~D. Hoffman, and Tony Jebara.
\newblock {Variational autoencoders for collaborative filtering}.
\newblock \emph{The Web Conference 2018 - Proceedings of the World Wide Web Conference, WWW 2018}, pp.\  689--698, 2 2018.
\newblock \doi{10.1145/3178876.3186150}.
\newblock URL \url{https://arxiv.org/abs/1802.05814}.

\bibitem[Mao et~al.(2021)Mao, Zhu, Wang, Dai, Dong, Xiao, and He]{Mao2021SimpleX:Filtering}
Kelong Mao, Jieming Zhu, Jinpeng Wang, Quanyu Dai, Zhenhua Dong, Xi~Xiao, and Xiuqiang He.
\newblock {SimpleX: A Simple and Strong Baseline for Collaborative Filtering}.
\newblock \emph{International Conference on Information and Knowledge Management, Proceedings}, pp.\  1243--1252, 9 2021.
\newblock \doi{10.1145/3459637.3482297}.
\newblock URL \url{https://arxiv.org/abs/2109.12613v2}.

\bibitem[Meng et~al.(2020)Meng, McCreadie, Macdonald, and Ounis]{Meng2020ExploringModels}
Zaiqiao Meng, Richard McCreadie, Craig Macdonald, and Iadh Ounis.
\newblock {Exploring Data Splitting Strategies for the Evaluation of Recommendation Models}.
\newblock \emph{RecSys 2020}, 7 2020.
\newblock \doi{10.48550/arxiv.2007.13237}.
\newblock URL \url{http://arxiv.org/abs/2007.13237}.

\bibitem[Page et~al.(1998)Page, Brin, Motwani, and Winograd]{Page1998TheWeb}
Lawrence Page, Sergey Brin, Rajeev Motwani, and Terry Winograd.
\newblock {The PageRank Citation Ranking: Bringing Order to the Web}.
\newblock \emph{World Wide Web Internet And Web Information Systems}, 54\penalty0 (1999-66), 1998.
\newblock ISSN 1752-0509.
\newblock \doi{10.1.1.31.1768}.

\bibitem[Patel et~al.(2022)Patel, Tolias, and Matas]{patel2022recall}
Yash Patel, Giorgos Tolias, and Ji{\v{r}}{\'\i} Matas.
\newblock Recall@ k surrogate loss with large batches and similarity mixup.
\newblock In \emph{{IEEE/CVF} CVPR}, 2022.

\bibitem[Perozzi et~al.(2014)Perozzi, Al-Rfou, and Skiena]{Perozzi2014DeepWalk:Representations}
Bryan Perozzi, Rami Al-Rfou, and Steven Skiena.
\newblock {DeepWalk: Online Learning of Social Representations}.
\newblock 3 2014.
\newblock \doi{10.1145/2623330.2623732}.
\newblock URL \url{http://arxiv.org/abs/1403.6652 http://dx.doi.org/10.1145/2623330.2623732}.

\bibitem[Pobrotyn \& Białobrzeski(2021)Pobrotyn and Białobrzeski]{https://doi.org/10.48550/arxiv.2102.07831}
Przemysław Pobrotyn and Radosław Białobrzeski.
\newblock {NeuralNDCG}: Direct optimisation of a ranking metric via differentiable relaxation of sorting, 2021.

\bibitem[Qin et~al.(2009)Qin, Liu, and Li]{Qin2009AGA}
Tao Qin, Tie-Yan Liu, and Hang Li.
\newblock A general approximation framework for direct optimization of information retrieval measures.
\newblock \emph{Information Retrieval}, 2009.

\bibitem[Ramzi et~al.(2021)Ramzi, Thome, Rambour, Audebert, and Bitot]{ramzi2021robust}
Elias Ramzi, Nicolas Thome, Cl{\'e}ment Rambour, Nicolas Audebert, and Xavier Bitot.
\newblock Robust and decomposable average precision for image retrieval.
\newblock \emph{NeurIPS}, 2021.

\bibitem[Ramzi et~al.(2022)Ramzi, Audebert, Thome, Rambour, and Bitot]{ramzi2022hierarchical}
Elias Ramzi, Nicolas Audebert, Nicolas Thome, Cl{\'e}ment Rambour, and Xavier Bitot.
\newblock Hierarchical average precision training for pertinent image retrieval.
\newblock In \emph{ECCV}, 2022.

\bibitem[Rendle et~al.(2012)Rendle, Freudenthaler, Gantner, and Schmidt-Thieme]{Rendle2012BPR:Feedback}
Steffen Rendle, Christoph Freudenthaler, Zeno Gantner, and Lars Schmidt-Thieme.
\newblock {BPR: Bayesian Personalized Ranking from Implicit Feedback}.
\newblock 5 2012.
\newblock URL \url{http://arxiv.org/abs/1205.2618}.

\bibitem[Revaud et~al.(2019)Revaud, Almaz{\'a}n, Rezende, and Souza]{revaud2019learning}
Jerome Revaud, Jon Almaz{\'a}n, Rafael~S Rezende, and Cesar Roberto~de Souza.
\newblock Learning with average precision: Training image retrieval with a listwise loss.
\newblock In \emph{CVPR}, 2019.

\bibitem[Rossi et~al.(2020)Rossi, Chamberlain, Frasca, Eynard, Monti, and Bronstein]{Rossi2020TGN:Graphs}
Emanuele Rossi, Ben Chamberlain, Fabrizio Frasca, Davide Eynard, Federico Monti, and Michael Bronstein.
\newblock {TGN: Temporal Graph Networks for Deep Learning on Dynamic Graphs}.
\newblock \emph{ArXiv 2020}, 6 2020.
\newblock URL \url{https://arxiv.org/abs/2006.10637}.

\bibitem[Sun et~al.(2020)Sun, Zhang, Guo, Guo, Tang, He, Ma, and Coates]{Sun2020NeighborRecommendation}
Jianing Sun, Yingxue Zhang, Wei Guo, Huifeng Guo, Ruiming Tang, Xiuqiang He, Chen Ma, and Mark Coates.
\newblock {Neighbor Interaction Aware Graph Convolution Networks for Recommendation}.
\newblock \emph{SIGIR 2020 - Proceedings of the 43rd International ACM SIGIR Conference on Research and Development in Information Retrieval}, pp.\  1289--1298, 7 2020.
\newblock \doi{10.1145/3397271.3401123}.
\newblock URL \url{https://dl.acm.org/doi/10.1145/3397271.3401123}.

\bibitem[Tang et~al.(2015)Tang, Qu, Wang, Zhang, Yan, and Mei]{Tang2015LINE:Embedding}
Jian Tang, Meng Qu, Mingzhe Wang, Ming Zhang, Jun Yan, and Qiaozhu Mei.
\newblock {LINE: Large-scale Information Network Embedding}.
\newblock \emph{WWW 2015 - Proceedings of the 24th International Conference on World Wide Web}, pp.\  1067--1077, 3 2015.
\newblock \doi{10.1145/2736277.2741093}.
\newblock URL \url{http://arxiv.org/abs/1503.03578 http://dx.doi.org/10.1145/2736277.2741093}.

\bibitem[Taylor et~al.(2008)Taylor, Guiver, Robertson, and Minka]{softrank}
Michael Taylor, John Guiver, Stephen Robertson, and Tom Minka.
\newblock Softrank: Optimizing non-smooth rank metrics.
\newblock In \emph{WSDM}, 2008.

\bibitem[Veli{\v{c}}kovi{\'{c}} et~al.(2017)Veli{\v{c}}kovi{\'{c}}, Cucurull, Casanova, Romero, Li{\`{o}}, and Bengio]{Velickovic2017GraphNetworks}
Petar Veli{\v{c}}kovi{\'{c}}, Guillem Cucurull, Arantxa Casanova, Adriana Romero, Pietro Li{\`{o}}, and Yoshua Bengio.
\newblock {Graph Attention Networks}.
\newblock 10 2017.
\newblock URL \url{http://arxiv.org/abs/1710.10903}.

\bibitem[Wang et~al.(2019)Wang, He, Wang, Feng, and Chua]{Wang2019NeuralFiltering}
Xiang Wang, Xiangnan He, Meng Wang, Fuli Feng, and Tat-Seng Chua.
\newblock {Neural Graph Collaborative Filtering}.
\newblock 5 2019.
\newblock \doi{10.1145/3331184.3331267}.
\newblock URL \url{https://arxiv.org/abs/1905.08108}.

\bibitem[Wang et~al.(2020)Wang, Jin, Zhang, He, Xu, and Chua]{Wang2020DisentangledFiltering}
Xiang Wang, Hongye Jin, An~Zhang, Xiangnan He, Tong Xu, and Tat-Seng Chua.
\newblock {Disentangled Graph Collaborative Filtering}.
\newblock \emph{SIGIR 2020}, 7 2020.
\newblock \doi{10.1145/3397271.3401137}.
\newblock URL \url{https://arxiv.org/abs/2007.01764}.

\bibitem[Wu et~al.(2021)Wu, Wang, Feng, He, Chen, Lian, and Xie]{Wu2021Self-supervisedRecommendation}
Jiancan Wu, Xiang Wang, Fuli Feng, Xiangnan He, Liang Chen, Jianxun Lian, and Xing Xie.
\newblock {Self-supervised Graph Learning for Recommendation}.
\newblock \emph{SIGIR 2021 - Proceedings of the 44th International ACM SIGIR Conference on Research and Development in Information Retrieval}, pp.\  726--735, 10 2021.
\newblock \doi{10.1145/3404835.3462862}.
\newblock URL \url{https://arxiv.org/abs/2010.10783}.

\bibitem[Xu et~al.(2018)Xu, Hu, Leskovec, and Jegelka]{Xu2018HowNetworks}
Keyulu Xu, Weihua Hu, Jure Leskovec, and Stefanie Jegelka.
\newblock {How Powerful are Graph Neural Networks?}
\newblock 10 2018.
\newblock URL \url{https://arxiv.org/abs/1810.00826}.

\bibitem[Ying et~al.(2018)Ying, He, Chen, Eksombatchai, Hamilton, and Leskovec]{Ying2018GraphPinSage}
Rex Ying, Ruining He, Kaifeng Chen, Pong Eksombatchai, William~L. Hamilton, and Jure Leskovec.
\newblock {Graph Convolutional Neural Networks for Web-Scale Recommender Systems (PinSage)}.
\newblock 6 2018.
\newblock \doi{10.1145/3219819.3219890}.
\newblock URL \url{http://arxiv.org/abs/1806.01973 http://dx.doi.org/10.1145/3219819.3219890}.

\bibitem[Yu et~al.(2022)Yu, Xia, Chen, Cui, Quoc, Hung, and Yin]{Yu2022XSimGCL:Recommendation}
Junliang Yu, Xin Xia, Tong Chen, Lizhen Cui, Nguyen Quoc, Viet Hung, and Hongzhi Yin.
\newblock {XSimGCL: Towards Extremely Simple Graph Contrastive Learning for Recommendation}.
\newblock 9 2022.
\newblock URL \url{https://arxiv.org/abs/2209.02544v4}.

\bibitem[Yue et~al.(2007)Yue, Finley, Radlinski, and Joachims]{Yue:2007}
Yisong Yue, Thomas Finley, Filip Radlinski, and Thorsten Joachims.
\newblock A support vector method for optimizing average precision.
\newblock In \emph{ACM SIGIR}, 2007.

\bibitem[Zhang et~al.(2010)Zhang, Guan, Sun, Liu, and Kong]{5609830}
Bangzuo Zhang, Yu~Guan, Haichao Sun, Qingchao Liu, and Jun Kong.
\newblock Survey of user behaviors as implicit feedback.
\newblock In \emph{2010 International Conference on Computer, Mechatronics, Control and Electronic Engineering}, volume~6, pp.\  345--348, 2010.
\newblock \doi{10.1109/CMCE.2010.5609830}.

\end{thebibliography}
\bibliographystyle{tmlr}

\appendix
\section{Method}\label{sec:sup_method}
In this section we define several ranking-based metrics used to evaluate recommender systems in~\cref{sec:sup_metric_definition}. In ~\cref{sec:sup_loss_definition} we show how to use the rank approximation of Eq. (1) of main paper to define other differentiable ranking-based losses.

\subsection{Metric definition}\label{sec:sup_metric_definition}

We remind here the definition of other well-known ranking-based metrics, used to evaluate recommender systems:
\begin{align}
    & R@k = \frac{\text{\# of positive in the top-$k$}}{\min(|\mathcal{V}^+|, k)} = \frac{1}{\min(|\mathcal{V}^+|, k)} \sum_{p\in\mathcal{V}^+} H(k - \rank(p)) \label{eq:sup_def_ratk} \\
    & \AP = \frac{1}{|\mathcal{V}^+|} \sum_{p\in\mathcal{V}^+} \frac{\rank^+(p)}{\rank(p)}, \text{ with } \rank^+(p) = \sum_{j\in\mathcal{V}^+} H(s_j-s_p) \label{eq:sup_def_ap}
\end{align}

\subsection{Other ranking based losses}\label{sec:sup_loss_definition}

In this section using the differentiable rank approximation of Eq. (1) of the main paper we define other differentiable ranking-based losses. In~\cref{sec:sup_model_analysis} we show the choice of the metric used during training impacts the evaluation metric.

The differentiable Average Precision loss is defined as follows (\cite{Qin2009AGA,brown2020smooth}):
\begin{equation}
    \label{eq:def_ap}
    \mathcal{L}_{\text{ITEM}}^{\text{AP}} = 1 - \frac{1}{|\mathcal{V}^+|} \sum_{p\in\mathcal{V}^+} \frac{\rank_s^+(p)}{\rank_s(p)}, \text{ with $\rank_s^+(p, \tau) = 1 + \sum_{j\in\mathbf{\mathcal{V}^+}} \sigma(s_j - s_p, \tau)$}
\end{equation}
Note that we use the approximation of Eq. (1) of the main paper to approximate both rank and $\rank^+$, using the same temperature $\tau$ parameter.

Finally, \cite{patel2022recall} defines an approximation the R@k as follows:
\begin{equation}\label{eq:def_recall}
    \mathcal{L}_{\text{ITEM}}^{R@k} = 1 - \frac{1}{|\mathcal{K}|} \sum_{k\in \mathcal{K}} \frac{1}{\min(|\mathcal{V}^+|, k)} \sum_{p\in\mathcal{V}^+} \sigma(k - \rank_s(p), \tau^*)
\end{equation}
In (\cite{patel2022recall}) the loss uses different level of recalls, \ie for $k$ in $\mathcal{K}$, it is necessary to provide enough gradient signal to all positive items. To train $\mathcal{L}_{\text{ITEM}}^{R@k}$, it is also necessary to approximate a second time the Heaviside function, using a sigmoid with temperature factor $\tau^*$.
Note that one other interesting property of both $\mathcal{L}_{\text{ITEM}}$ and $\mathcal{L}_{\text{ITEM}}^{\text{AP}}$ is that they only require to select one hyper-parameter, $\tau$.

\section{Experiments}

\subsection{Datasets}
\textcolor{black}{We evaluate our approach in the implicit feedback recommendation task \cite{5609830}, simulating real-world scenarios where explicit feedback is too costly. The relationships are binary: if a user interacts with an item, it indicates their appreciation, represented by a link in the graph.
We validate our method on four recommendation datasets. We use the latest small MovieLens-100k  dataset (\cite{harper2015movielens}) in the inductive settings. It features relations between users and movies, where users give ratings to different movies. For implicit feedback, we keep relations with ratings of 3 or higher. We use the 2018 edition of the dataset from the Yelp challenge (\cite{asghar2016yelp}). Rated items are bars and restaurants, we follow (\cite{Wang2019NeuralFiltering,He2020LightGCN:Recommendation}) to get implicit feedback. We also evaluate ITEM on the Amazon-Book dataset from Amazon-review, we use the same pre-processing as NGCF (\cite{Wang2019NeuralFiltering}) and LightGCN (\cite{He2020LightGCN:Recommendation}).
For the three datasets we apply a pre-processing to make sure that we have the same items for the training and for the evaluation. 
We follow \cite{Wang2019NeuralFiltering} and only keep users that have at least 10 interactions. 
Tab. 1 in main paper presents the datasets statistics.} 
\newpage
\subsection{Implementation details} \textcolor{black}{All our models are trained using the Adam optimizer, with learning rate in \{0.01, 0.001\}. On the inductive setting, we use $\tau=1.0$, and $\tau=1.5$ on the transductive setting. We use sum-pooling and embeddings of dimension 200 for the inductive setting on Yelp and MovieLens-100k and embeddings of dimension 64 for Amazon-Book, sum-pooling -- except for LightGCN (\cite{He2020LightGCN:Recommendation}) with mean-pooling -- and embeddings of dimension 64 for the transductive setting. On both settings, we sample 5 positives and 200 negatives for ITEM for Yelp2018, MovieLens-100k and MovieLens-1M and 600 negatives for AmazonBook. We use, on both protocols, batch sizes of 512 for MovieLens-100k, MovieLens-1M, and 2048 for Yelp-2018 and Amazon-book. For the inductive setting, we found experimentally that learning user embeddings during training is harmful for the generalization of the GNNs to new users, so for training and evaluation users' embeddings are inferred using message passing only.
For GIN (\cite{Xu2018HowNetworks}), GAT (\cite{Velickovic2017GraphNetworks}) and GCN (\cite{Kipf2016Semi-SupervisedNetworks}) we use the framework Pytorch Geometric (\cite{Fey/Lenssen/2019}) to implement the models. }

\subsection{Ablation studies}

\begin{table*}[ht]
    \setlength\tabcolsep{1pt}
    \setlength\extrarowheight{4pt}
    \caption{Comparison of our ranking-based loss ITEM vs the BPR loss (\cite{Rendle2012BPR:Feedback}), using different GNN architecture on 3 \textit{inductive} benchmarks. For each architecture best results is \textbf{bold}, best overall results \underline{underlined}.}
    \label{tab:sup_main_inductive} 
    \centering
    \begin{tabularx}{\textwidth}{l l YY | YY | YY }
        \toprule
         & \multirow{2}{*}{Method} & \multicolumn{2}{c|}{MovieLens-100k} & \multicolumn{2}{c|}{Yelp-2018} & \multicolumn{2}{c}{Amazon-book} \\
        \cmidrule{3-8}
         &&   R@20 &   NDCG@20 &  R@20 &   NDCG@20 &  R@20 &  NDCG@20 \\
         \midrule

         \multirow{3}{*}{\rotatebox[origin=c]{90}{GCN}}
         & BPR & 28.74 & 27.68 & 7.34 & 5.76 & 8.85 & 7.61\\

         & $\llargo$ & \textbf{32.24}  & \textbf{31.44} & \textbf{10.06} & \textbf{8.35} & \textbf{10.23} & \textbf{8.82}\\
         &  \small \%Improv. & \small  {+12.2\%} & \small  {+13.5\%} & \small  {+37.1\%} & \small  {+45\%} & \small  {+15.6\%} & \small  {+15.9\%} \\
         \midrule
         \multirow{3}{*}{\rotatebox[origin=c]{90}{GIN~}}
         & BPR & 29.71 & 27.58 & 8.25 & 6.74 & 9.62 & 8.05 \\

         & $\llargo$ & \textbf{32.00} & \textbf{29.87} & \textbf{10.57} & \textbf{8.79} & \underline{\textbf{13.05}} & \underline{\textbf{11.59}}\\
         & \small \%Improv. & \small  {+7.7\%} & \small  {+8.3\%} & \small  {+28.5\%} & \small  {+30.4\%} & \small  {+35.7\%} & \small  {+44.0\%} \\
          \midrule
         \multirow{3}{*}{\rotatebox[origin=c]{90}{GAT~}}
         & BPR & 31.01 & 28.92 & 9.04 & 7.32 & 9.88 & 8.17\\

          & $\llargo$ & \textbf{32.04} & \textbf{30.54} & \underline{\textbf{11.43}} & \underline{\textbf{9.58}} & \textbf{11.70} & \textbf{10.00} \\
         & \small \%Improv. & \small  {+3.3\%} & \small  {+5.6\%} & \small  {+26.4\%} & \small  {+30.9\%} & \small  {+18.4\%} & \small  {+22.4\%} \\
         \midrule
         \multirow{3}{*}{\rotatebox[origin=c]{90}{LGCN~}}
         & BPR & 30.79 & 29.73 & 7.88 & 6.34 & 9.56 & 8.02 \\
         & $\llargo$ & \underline{\textbf{33.13}} & \underline{\textbf{32.07}} & \textbf{9.88} & \textbf{8.32} & \textbf{10.66} & \textbf{9.52} \\
         &  \small \%Improv. & \small  {+7.6\%} & \small  {+7.9\%} & \small  {+25.4\%} & \small  {+31.2\%} & \small  {+11.5\%} & \small  {+18.7\%}\\
         \bottomrule
    \end{tabularx}
\end{table*}

\subsubsection{Inductive loss comparison} In~\cref{tab:sup_main_inductive} we compare in the same settings the BPR loss (\cite{Rendle2012BPR:Feedback}) and our proposed ranking-based loss~ITEM. We show that on the three inductive benchmarks, and across all four considered architectures, ITEM outperforms the BPR loss. On \textcolor{black}{MovieLens-100k} and with the best performing architecture, LightGCN (\cite{He2020LightGCN:Recommendation}), our loss outperforms the BPR loss with \textcolor{black}{+2.34 R@20 and +2.34 NDCG@20}. \textcolor{black}{The relative improvements are also significant, ranging from +3.3\% NDCG@20 for GCN, up to 12.2\% NDCG@20 for GCN}. \textcolor{black}{On Yelp-2018 which is a large scale dataset, GAT is the best performing architecture. ITEM outperforms the BPR loss by +2.39 R@20 and +2.26 NDCG@20.} \textcolor{black}{We can point out the considerable relative improvements on Yelp-2018, which are always larger than $25\% $ and reaching $45\%$ over GCN for NDCG@20.} Finally on Amazon-book the best overall results are obtained with GIN using ITEM. It outperforms BPR and GIN by +3.4 R@20 and 3.5 NDCG@20 which are huge relative improvements on Amazon-book our most large scale dataset. Note that accros the three considered datasets different GNN architectures work best, LightGCN (\cite{He2020LightGCN:Recommendation}) on MovieLens-100k, GAT (\cite{Velickovic2017GraphNetworks}) on Yelp-2018 and GIN (\cite{Xu2018HowNetworks}) on Amazon-book.
Overall, \cref{tab:sup_main_inductive} -- similarly as Tab. 4 in the main paper -- shows the interest of optimizing a ranking-based loss, $\llargo$, rather than the BPR loss, \ie a \textit{proxy} loss.

\subsubsection{Negative sampling} In~\cref{tab:sup_sampling_compa} we show the impact of our PPR negative sampling strategy in the inductive setting. We show that on the three inductive benchmarks using our sampling strategy boost the performances of our trained GNN model over the random negative sampling (RNS), \eg +0.66 R@20 and +0.4 NDCG@20 on Yelp-2018.

\begin{table*}[ht]
    \setlength\tabcolsep{0.3pt}
    \setlength\extrarowheight{1pt}
    \caption{Comparison of Random Negative Sampling (RNS) vs the sampling method in ITEM (PPR). We use a LightGCN (\cite{He2020LightGCN:Recommendation}) backbone for the three \textit{\textbf{inductive}} benchmarks.}
    \label{tab:sup_sampling_compa} 
    \centering
    \begin{tabularx}{\textwidth}{lc YY | YY | YY }
        \toprule
         \multirow{2}{*}{Loss} & \multirow{2}{*}{Sampling} & \multicolumn{2}{c|}{MovieLens-100k} & \multicolumn{2}{c|}{Yelp-2018} & \multicolumn{2}{c}{Amazon-book} \\
         && R@20 & NDCG@20 & R@20 & NDCG@20 & R@20 & NDCG@20 \\
         \midrule
         \multirow{2}{*}{$\llargo$} & RNS & {{33.13}} & {{32.07}} & {9.88} & {8.32} & 10.66 & 9.52 \\
         &  PPR (ours) & \textbf{33.84} & \textbf{32.63} & \textbf{10.54} & \textbf{8.70} & \textbf{11.03} & \textbf{9.89}  \\
         \bottomrule
    \end{tabularx}
\end{table*}
\subsubsection{Cold-start evaluation}
\textcolor{black}{In \cref{tab:cold_start}, we evaluate our model in a cold-start setting with only a small proportion of training data available for each user (10\%, 20\%, 50\%). Our method, which uses an approximation of NDCG and PPR-based negative sampling, shows significant robustness compared to the baseline LightGCN with BPR loss. }
\begin{table*}[!h]
    \setlength\tabcolsep{0.3pt}
    \setlength\extrarowheight{1pt}
    \caption{Cold-start evaluation in a transductive settings on MovieLens-1M with different amount of training data (\%).}
    \label{tab:cold_start} 
    \centering
    \begin{tabularx}{\textwidth}{l YY | YY | YY }
        \toprule
         \multirow{2}{*}{Method}  & \multicolumn{2}{c|}{10\%} & \multicolumn{2}{c|}{20\%} & \multicolumn{2}{c}{50\%} \\
         & R@20 & NDCG@20 & R@20 & NDCG@20 & R@20 & NDCG@20 \\
         \midrule
         LGCN-BPR & 15.88 & 15.43 & 20.05 & 19.18 & 24.19 &23.36 \\
         \rowcolor[gray]{.95} ITEM & \textbf{17.11} & \textbf{16.89} & \textbf{22.37} &\textbf{ 21.68} &\textbf{ 29.03} & \textbf{28.20}\\
         \bottomrule
    \end{tabularx}
\end{table*}

\subsection{Model analysis}\label{sec:sup_model_analysis}

\subsubsection{Metric optimization} In~\cref{tab:loss_compa}, we compare the benefits of optimizing different ranking-based losses on MovieLens-100k (inductive) using a LightGCN model. Specifically, we compare the optimization of AP, R@k and NDCG, and include the BPR loss as a baseline. First we can note that on each metric all our ranking-based losses outperform the BPR loss. We can see that, for each metric, using its smooth approximation to optimize a GNN during training yields a higher score on this target metric. $\LAPs$ yields the best score of 19.05 AP, $\LRks$ yields the best score of 33.37 R@20 (outperforming the results reported in~\cref{tab:sup_main_inductive}). Finally $\LNDCGs$ yields the best score for both NDCG and NDCG@20 of 53.55 and 32.07 respectively.

\begin{table*}[ht]
    \setlength\tabcolsep{0.3pt}
    \setlength\extrarowheight{1pt}
    \caption{Performances of different ranking-based losses on the MovieLens-100k inductive benchmark, and the BPR loss~\cite{Rendle2012BPR:Feedback} baseline. The model used is LightGCN~\cite{He2020LightGCN:Recommendation}.}
    \label{tab:loss_compa} 
    \centering
    \begin{tabularx}{\textwidth}{l YYYY }
        \toprule
         Loss  & AP &  NDCG & R@20 &  NDCG@20 \\
         \midrule
         BPR &16.56&50.68& 29.59 & 28.24 \\
         \midrule
         $\mathcal{L}_{\text{ITEM}}^{\text{AP}}$ & \textbf{19.05} & 53.48 & 32.61 & 31.91 \\
         $\mathcal{L}_{\text{ITEM}}^{R@k}$ & 18.57 & 52.93 & \textbf{33.37} & 31.82 \\
         $\mathcal{L}_{\text{ITEM}}$ & {18.94}&\textbf{53.55}& {33.13} & \textbf{32.07} \\
         \bottomrule
    \end{tabularx}
\end{table*}

\subsection{Qualitative Results}\label{sec:supp_qualitative_results}

\begin{figure}[ht]
    \centering
    \includegraphics[width=\textwidth]{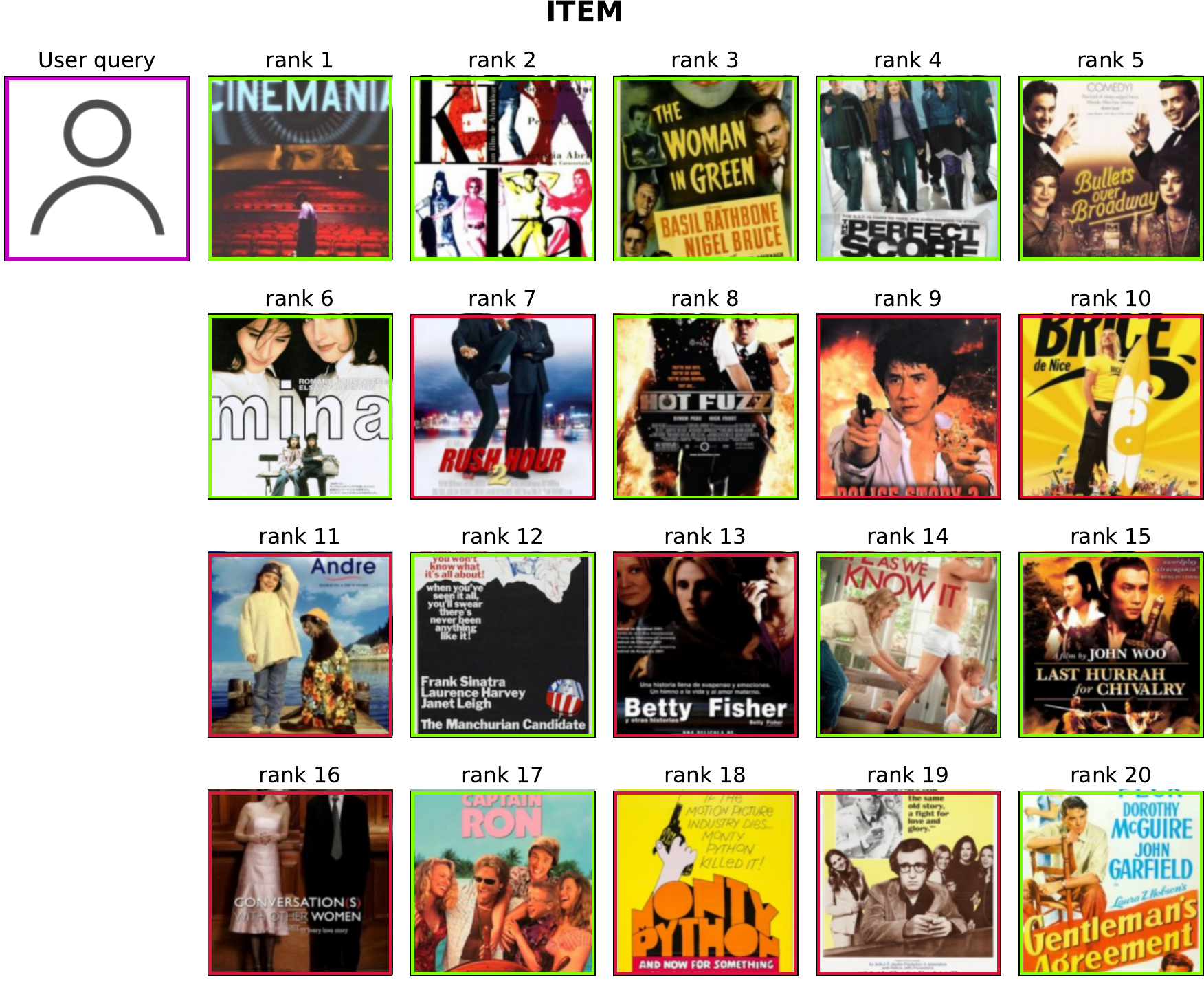}
    \caption{Top-20 results on MovieLens-100k (inductive) when using LightGCN and \textbf{ITEM}.}
    \label{fig:sup_qual_results_ours}
\end{figure}

\begin{figure}[ht]
    \centering
    \includegraphics[width=\textwidth]{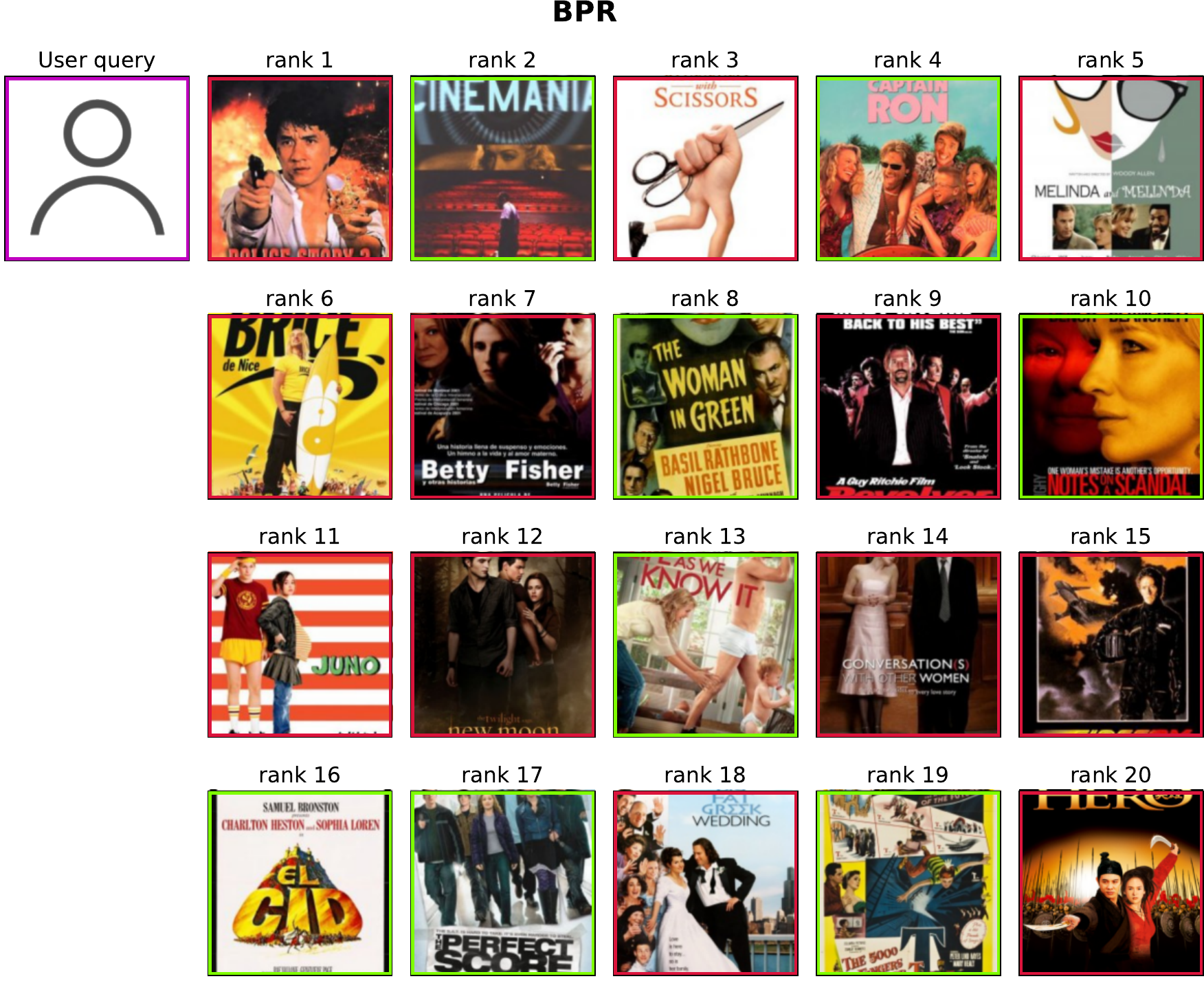}
    \caption{Top-20 results on MovieLens-100k (inductive) when using LightGCN and BPR loss.}
    \label{fig:sup_qual_results_bpr}
\end{figure}

We display on~\cref{fig:sup_qual_results_ours} the top-20 items retrieved when using ITEM on MovieLens-100k, and on~\cref{fig:sup_qual_results_bpr} the top-20 items retrieved when using the baseline BPR loss (\cite{Rendle2012BPR:Feedback}). Both models are LightGCN (\cite{He2020LightGCN:Recommendation}). We can observe qualitatively that ITEM brings more positive results, and leads to a better ranking than the BPR loss. 

\end{document}